\documentclass[twocolumn]{aastex6}
\usepackage{graphicx}
\usepackage{amssymb,amsfonts,amsmath,amstext,amsgen,amsopn,amsxtra,indentfirst,times}

\begin{document}

\title{Information Content of JWST-NIRSpec Transmission Spectra of Warm Neptunes}

\author{Andrea Guzm\'{a}n-Mesa\altaffilmark{1}}
\author{Daniel Kitzmann\altaffilmark{1}}
\author{Chloe Fisher\altaffilmark{1,6}}
\author{Adam J. Burgasser\altaffilmark{2,7}}
\author{H. Jens Hoeijmakers\altaffilmark{1,3}}
\author{Pablo M\'arquez-Neila\altaffilmark{1,4}}
\author{Simon L. Grimm\altaffilmark{1}}
\author{Avi M. Mandell\altaffilmark{5}}
\author{Raphael Sznitman\altaffilmark{4}}
\author{Kevin Heng\altaffilmark{1,8}}
\altaffiltext{1}{University of Bern, Center for Space and Habitability, Gesellschaftsstrasse 6, CH-3012, Bern, Switzerland.  Emails: andrea.guzmanmesa@space.unibe.ch, kevin.heng@csh.unibe.ch}
\altaffiltext{2}{Center for Astrophysics and Space Science, University of California San Diego, La Jolla, CA 92093, U.S.A.}
\altaffiltext{3}{Observatoire astronomique de l'Université de Gen\`{e}ve, 51 chemin des Maillettes, 1290 Versoix, Switzerland}
\altaffiltext{4}{University of Bern, ARTORG Center for Biomedical Engineering, Murtenstrasse 50, CH-3008, Bern, Switzerland}
\altaffiltext{5}{Solar System Exploration Division, NASA Goddard Space Flight Center, Greenbelt, MD 20771, U.S.A.}
\altaffiltext{6}{University of Bern International 2021 Ph.D Fellowship}
\altaffiltext{7}{On sabbatical at the Center for Space and Habitability in Fall 2019}
\altaffiltext{8}{University of Warwick, Department of Physics, Astronomy \& Astrophysics Group, Coventry CV4 7AL, U.K.}

\begin{abstract}
Warm Neptunes offer a rich opportunity for understanding exo-atmospheric chemistry.  With the upcoming \textit{James Webb Space Telescope} (JWST), there is a need to elucidate the balance between investments in telescope time versus scientific yield.  We use the supervised machine learning method of the random forest to perform an information content analysis on a 11-parameter model of transmission spectra from the various NIRSpec modes.  The three bluest medium-resolution NIRSpec modes (0.7--1.27 $\mu$m, 0.97--1.84 $\mu$m, 1.66--3.07 $\mu$m) are insensitive to the presence of CO.  The reddest medium-resolution mode (2.87--5.10 $\mu$m) is sensitive to all of the molecules assumed in our model: CO, CO$_2$, CH$_4$, C$_2$H$_2$, H$_2$O, HCN and NH$_3$.  It competes effectively with the three bluest modes on the information encoded on cloud abundance and particle size.  It is also competitive with the low-resolution prism mode (0.6--5.3 $\mu$m) on the inference of every parameter except for the temperature and ammonia abundance.  We recommend astronomers to use the reddest medium-resolution NIRSpec mode for studying the atmospheric chemistry of 800--1200 K warm Neptunes; its corresponding high-resolution counterpart offers diminishing returns.  We compare our findings to previous JWST information content analyses that favor the blue orders, and suggest that the reliance on chemical equilibrium could lead to biased outcomes if this assumption does not apply. A simple, pressure-independent diagnostic for identifying chemical disequilibrium is proposed based on measuring the abundances of H$_2$O, CO and CO$_2$.
\end{abstract}


\section{Introduction}
\label{sect:intro}

With the much anticipated launch of the \textit{James Webb Space Telescope} (JWST) in 2021 (and Cycle 1 proposals due in 2020), the exoplanet community is studying the balance between investments of telescope time and scientific yield \citep{beichman14,barstow15,greene16,howe17,bl17}.  Both the Guaranteed Time Observations as well as the Early Release Science programs are designed to gain an understanding of systematics and data reduction strategies \citep{stevenson16,bean18,kil18} and will provide the first opportunities to obtain JWST transit spectroscopy data over a wide range of infrared wavelengths for many of the best-known transiting exoplanets.  

\subsection{Motivation I: anticipated chemical diversity of warm Neptunes}

One of the unexpected outcomes of the Kepler mission is that $\sim 1000$ K sub-Neptune- to Neptune-sized exoplanets on short-period orbits are common (e.g., \citealt{petigura13,crossfield16}), which we will collectively term ``warm Neptunes" in the current study.  Their bulk densities indicate the presence of a hydrogen- and/or helium-dominated atmosphere.  The \textit{Transiting Exoplanet Survey Satellite} (TESS) is discovering warm Neptunes orbiting bright stars (e.g., \citealt{drago19,esposito19,quinn19,trifonov19}). With no example in our Solar System, a deeper understanding of the properties of warm Neptunes is expected to shed light on exoplanet formation processes.  The complete chemical inventory of their atmospheres is currently unknown and it is expected that JWST spectra will allow the exoplanet community to make significant progress on this question.

Across a temperature range of 800--1200 K, warm Neptunes are theoretically predicted to exhibit remarkable chemical diversity with water (H$_2$O), methane (CH$_4$), carbon dioxide (CO$_2$) and carbon monoxide (CO) having a wide range of volume mixing ratios as the elemental abundance of carbon (C/H) and the carbon-to-oxygen ratio (C/O) vary \citep{moses13}.  At $\sim 1000$ K, equilibrium chemistry predicts a transition from CH$_4$- to CO-dominated atmospheres toward higher temperatures (e.g., \citealt{moses11,madhu12,venot14,ht16}).  However, 800--1200 K is also the temperature range where the assumption of chemical equilibrium breaks down, because the chemical and dynamical timescales become comparable and photochemistry may not be negated by high temperatures.  For example, \cite{ms11} find tentative evidence for the over-abundance of CO (compared to expectations from chemical equilibrium) in the warm Neptune GJ 436b, which has an equilibrium temperature of about $649 \pm 60$ K \citep{torres08}; see also \citep{morley17}.  In our own Jupiter, the over-abundance of CO was interpreted as a sign of disequilibrium chemistry due to atmospheric mixing \citep{pb77}.  Similarly, \cite{opp98} detected an excess of CO in the brown dwarf Gliese 229B.

For all of these reasons, warm Neptunes with atmospheric temperatures in the range of 800--1200 K are the next frontier in understanding atmospheric chemistry from transmission spectroscopy.

\subsection{Motivation II: accuracy of constraining elemental abundances and C/O}

The key controlling parameters of atmospheric chemistry are the set of elemental abundances (mainly C/H, O/H, N/H) and C/O (e.g., \citealt{bs99,madhu12,ht16}).  Atmospheric mixing and photolysis act to complicate the translation between the elemental and molecular abundances (e.g., \citealt{moses11,tsai17}).  As already noted by \cite{line13} and \cite{greene16}, the best approach for inferring the elemental abundances and C/O from spectra is to directly retrieve the abundances of the major carbon, oxygen and nitrogen molecular carriers,
\begin{equation}
\begin{split}
\mbox{C/H} &= \frac{X_{\rm CO}+X_{\rm CO_2} + X_{\rm CH_4} + X_{\rm HCN} + 2X_{\rm C_2H_2}}{X_{\rm H}}, \\
\mbox{O/H} &= \frac{X_{\rm CO}+2X_{\rm CO_2} + X_{\rm H_2O}}{X_{\rm H}}, \\
\mbox{N/H} &= \frac{X_{\rm NH_3}+X_{\rm HCN}}{X_{\rm H}}, \\
\mbox{C/O} &= \frac{X_{\rm CO}+X_{\rm CO_2} + X_{\rm CH_4} + X_{\rm HCN} + 2X_{\rm C_2H_2}}{X_{\rm CO}+2X_{\rm CO_2} + X_{\rm H_2O}},
\end{split}
\label{eq:ratios}
\end{equation}
where $X_i$ are the volume mixing ratios of molecules and $X_{\rm H} = 2 X_{\rm H_2} + 4X_{\rm CH_4} + X_{\rm HCN} + 2X_{\rm C_2H_2} + 2X_{\rm H_2O} + 3X_{\rm NH_3}$.  In H$_2$-dominated atmospheres (as studied here), $X_{\rm H} \approx 2 X_{\rm H_2}$.  It is important to note that these are inferred quantities in the gas phase, which may differ from their bulk values due to condensation, e.g., sequestration of oxygen into olivine.  Only in extremely hot conditions, such as for ultra-hot Jupiters and main-sequence stars, may we reasonably assume that the photospheric and bulk elemental abundances are similar (e.g., \citealt{kitzmann18}).  \cite{line13} cautions that retrieving directly for the molecular abundances results in a non-uniform prior for C/O (see their Section 3.3).

In the current study, we consider 7 molecules.  CO and CH$_4$ are the major carbon carriers \citep{bs99} with CO$_2$ being a minor carbon carrier unless the metallicity is highly enriched (e.g., \citealt{moses13,hl16}).  H$_2$O and CO are major oxygen carriers \citep{bs99}.  Acetylene (C$_2$H$_2$) becomes non-negligible as C/O approaches unity (e.g., \citealt{moses11,madhu12,ht16}).  NH$_3$ competes with molecular nitrogen (N$_2$) as the major nitrogen carrier \citep{bs99}, while hydrogen cyanide (HCN) is an important link between the carbon and nitrogen reservoirs (e.g., \citealt{moses11}).  The accuracy of retrieving for the elemental abundances hinges on a spectrum having sufficient spectral resolution, signal-to-noise and wavelength coverage to accurately account for the molecules that are present in sufficient amounts.  If not all of the molecules are properly accounted for, it will lead to erroneous inferences about C/O.

A second approach is to assume chemical equilibrium and parametrise all of the molecular abundances by two numbers: C/O and the metallicity.  Chemical equilibrium is a \textit{local} approximation in the sense that each patch of atmosphere has no memory of its past and all of the molecular abundances may be completely determined once one has knowledge of the local temperature and pressure.  Metallicity has three definitions in the astronomical literature: stellar astrophysicists refer to the relative abundance of all elements heavier than helium \textit{by mass} (Section 3.12 of \citealt{asplund09}), observational spectroscopists refer to the elemental abundance of iron \textit{by number} (Section 4.2 of \citealt{asplund09}) and atmospheric chemists typically refer to the elemental abundance of a volatile element (e.g., carbon) by number \citep{moses13}.  In the third definition, it is usually assumed that the ratios of the elemental abundances are kept fixed to their solar values with the exception of C/H or O/H, which are allowed to be free parameters in order to allow for a variable C/O (e.g., \citealt{moses13,heng18,drummond19}).  In chemical equilibrium, knowledge of the abundance of a single carbon or oxygen carrier is sufficient to constrain C/H or O/H, respectively.  However, if chemical equilibrium is a poor assumption, then misleading conclusions will follow.  None of the atmospheres of Solar System bodies are well-described by chemical equilibrium.

One of the goals of the current study is to examine the relationship between the accuracy of retrieving for the elemental abundances and hence C/O.

\begin{table}[!ht]
\label{tab:modes}
\begin{center}
\caption{JWST NIRSpec Modes Considered}
\begin{tabular}{llccc}
\hline
Shorthand & Wavelengths ($\mu$m) & Configuration & Resolution  \\
\hline
L & 0.6--5.3 & PRISM/CLEAR & 100 \\
M1 & 0.7--1.27 & G140M/F070LP & 600 \\
M2 & 0.97--1.84 & G140M/F100LP & 1000 \\
M3 & 1.66--3.07 & G235M/F170LP & 1000 \\
M4 & 2.87--5.10 & G395M/F290LP & 1000 \\
H4 & 2.87--5.10 & G395H/F290LP & 2700 \\
\hline
\hline
\end{tabular}\\
\end{center}
\end{table}

\subsection{Motivation III: novel information content analysis approach, feasible for complex models}

\begin{figure*}%[!t]
\begin{center}
%\vspace{-0.2in}
\includegraphics[width=2\columnwidth]{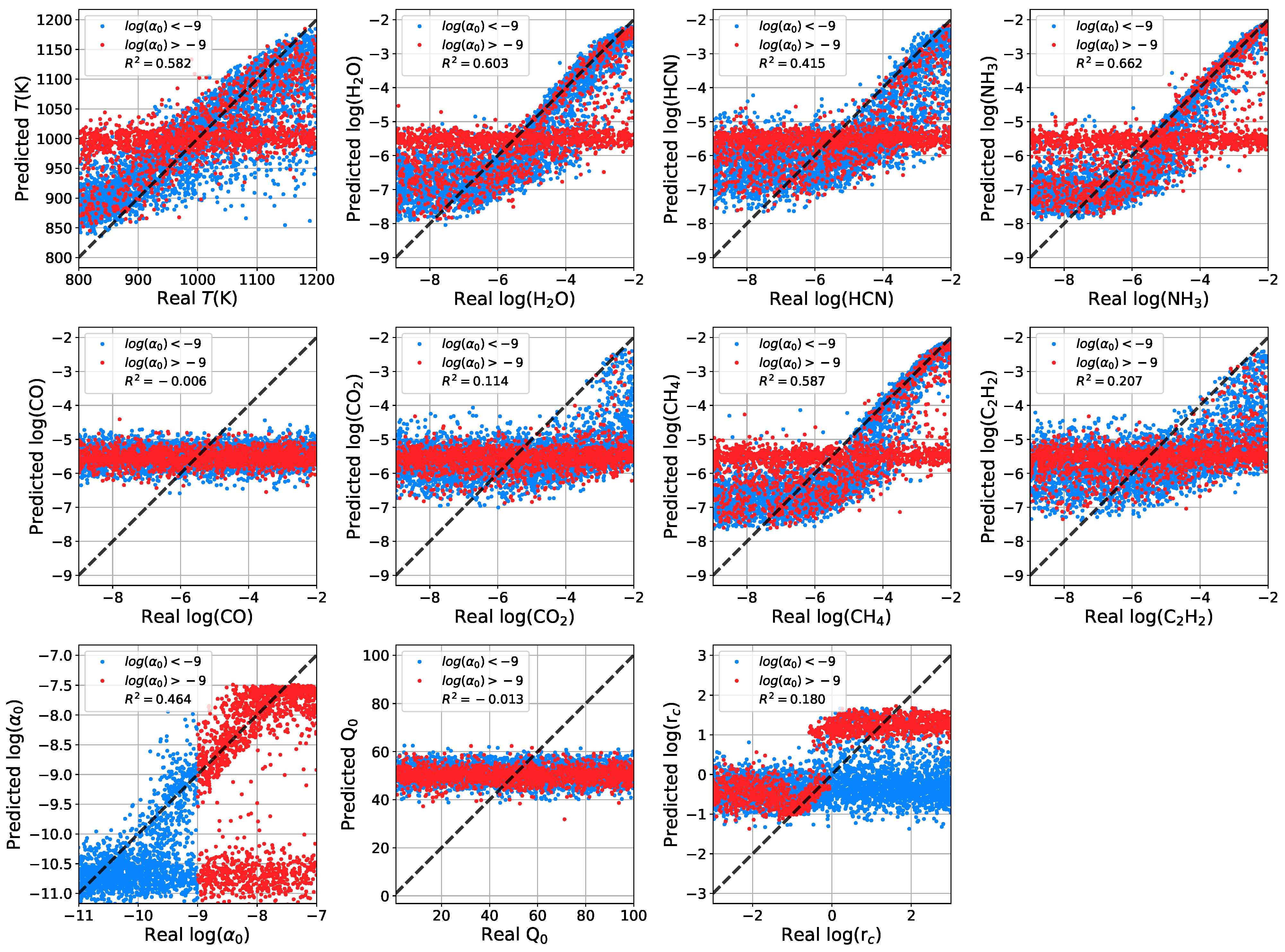}
\end{center}
\vspace{-0.1in}
\caption{Real versus predicted (RvP) values of the various parameters from a suite of 20,000 mock retrievals on HST-WFC3 transmission spectra using the random forest method.  For clarity (and with no loss of generality), we only show 5000 of these mock retrievals.  The stellar and exoplanetary parameters of GJ 436 and the warm Neptune GJ 436b, respectively, are assumed (see text).  The synthetic spectra are composed of 13 wavelength bins from 0.8--1.7 $\mu$m following \cite{k15} and to provide continuity with \cite{mn18}.  Each synthetic data point assumes an optimistic photon-limited uncertainty of 20 parts per million (ppm). The blue and red points correspond to cloudfree ($\alpha_0 < 10^{-9}$ cm$^{-1}$; see text for definition) and cloudy ($\alpha_0 > 10^{-9}$ cm$^{-1}$) models, respectively. Negative and positive values of the coefficient of determination (${\cal R}$) correspond to negative and positive correlations, respectively.}
%\vspace{-0.1in}
\label{fig:wfc3_rvp}
\end{figure*}

\begin{figure}%[!t]
\begin{center}
%\vspace{-0.2in}
\includegraphics[width=\columnwidth]{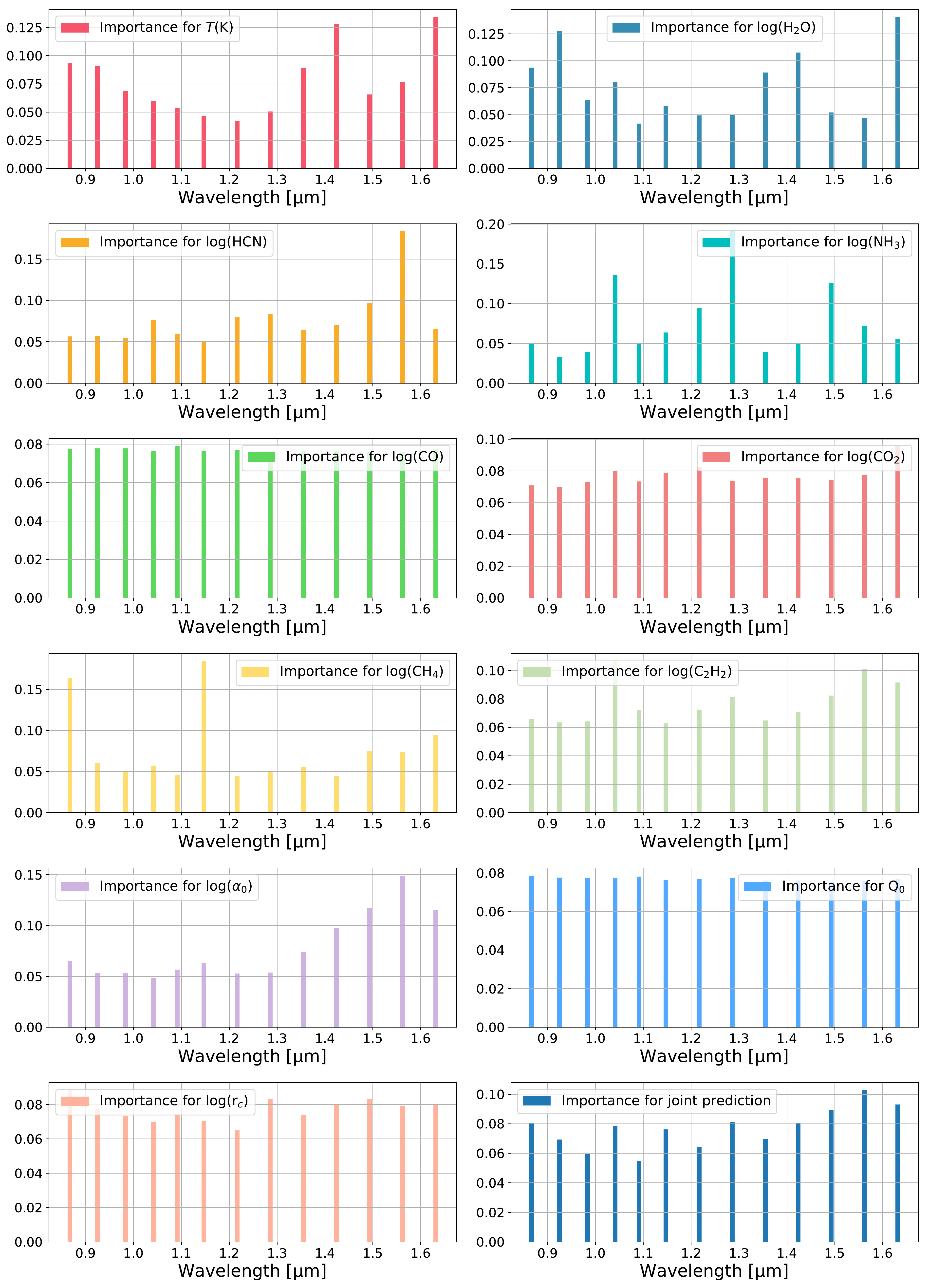}
\end{center}
\vspace{-0.1in}
\caption{Companion figure montage to Figure \ref{fig:wfc3_rvp}, which shows the ``feature importance" plots from the random forest retrieval analysis.  Each feature importance plot quantifies the relative importance of each data point in a HST-WFC3 transmission spectrum for determining the value of a given parameter.  The entries in each panel add up to unity.}
%\vspace{-0.1in}
\label{fig:wfc3_feature}
\end{figure}

Classical information content analysis is based on computing Jacobians, which are the derivatives of the model output (e.g., transit depth) with respect to the parameters.  See Section 2 of \cite{bl17} for a recent review.\footnote{Note that the treatment in \cite{bl17} requires the assumption of Gaussian probability distributions.}  Classical information content analysis is a time-consuming process.  For example, Section 3.1 of \cite{bl17} states, ``For each of these 84 combinations of planet types, we compute a \textit{separate} Jacobian" (these authors' emphasis).  \cite{howe17} introduces the use of ``mutual information" (see their Section 2), but remark how ``the difficulty with the use of mutual information is that it is computationally intensive, especially for the dense data sets produced by JWST."  For reasons of computational feasibility, \cite{howe17} adopted a simple three-parameter model that assumed an isothermal transit chord, gray clouds and a metallicity.\footnote{Presumably, this requires the assumption of chemical equilibrium, but \cite{howe17} do not explicitly state this beyond the following sentence: ``Most notably, the alkali metal lines and the CO bands grow much stronger with increasing temperature as the concentrations of these species in chemical equilibrium increase."}  \cite{bl17} assumed chemical equilibrium models described by the metallicity and C/O and a non-gray treatment of clouds.

In the current study, we adopt a qualitatively different approach to information content analysis.  Recently, \cite{mn18} demonstrated that the classical machine learning method of the ``random forest" \citep{ho98,breiman01,crimi11} may be adapted to perform atmospheric retrieval, as a complement to standard methods such as nested sampling \citep{skilling06,feroz08,feroz09,feroz13}.  \cite{fisher20} compares random forest retrieval to other methods (nested sampling, Bayesian neural networks).  There are three distinct advantages of random forest retrieval in terms of practical implementation.  First, it performs ``feature importance", which in the context of spectra means it is able to compute the relative importance of each data point for constraining each parameter of a chosen model used to interpret the spectra.  Second, it is able to easily perform large suites of mock retrievals in the form of ``real versus predicted" (RvP) plots \citep{mn18,fisher20,oreshenko20}.  Third, since the random forest may be trained on a pre-computed model grid of arbitrary sophistication, the obstacles of computational feasibility encountered by \cite{howe17} and \cite{bl17} may be overcome.  Instead of assuming chemical equilibrium, we allow each of our 7 molecules to take on a broad range of abundances and infer the elemental abundances and C/O from the retrieved abundances.

Examples of RvP plots are shown in Figure \ref{fig:wfc3_rvp}, where we perform a suite of 20,000 mock retrievals for Hubble Space Telescope (HST) Wide Field Camera 3 (WFC3) transmission spectra of the warm Neptune GJ 436b.  These RvP plots may be used to quantify the ability of a retrieval to accurately recover each parameter value of the model.  The random forest reports the \textit{mean} predicted values of the parameters in the RvP plots.  The figure of merit used is the ``coefficient of determination" (${\cal R}^2$), where ${\cal R}^2=0$ means zero predictability (zero correlation between the real versus predicted values of a parameter) and ${\cal R}^2=1$ means perfect predictability.  Model degeneracies will generally lower the value of ${\cal R}^2$ \citep{mn18,fisher20,oreshenko20}.  The RvP plots reproduce widely accepted knowledge in the exoplanet retrieval literature: WFC3 transmission spectra probe mainly H$_2$O, CH$_4$ and NH$_3$ with some sensitivity to HCN, but are insensitive to CO and CO$_2$.  Furthermore, while cloud particle radius and abundance may be retrieved, one is blind to the retrieval of cloud composition.  If CO and H$_2$O are present in comparable abundances, then the retrieval will only accurately infer the H$_2$O abundance, leading to an inaccurate estimate of C/O.

Figure \ref{fig:wfc3_feature} shows the accompanying feature importance plots.  Each feature importance plot quantifies the relative importance of each of the 13 data points in the WFC3 transmission spectrum for determining the value of a parameter.  For example, the two data points near 1.4 $\mu$m constrain the water abundance, which matches our intuition of a water feature being present at these wavelengths.  The bluest data points constrain the cloud abundance and particle radius.  When the feature importance is about equal for all 13 data points (e.g., for CO), it often indicates a lack of sensitivity to a given parameter, which can only be confirmed by cross-matching with the ${\cal R}^2 \approx 0$ value from the RvP plot (Figure \ref{fig:wfc3_rvp}).

In the current study, we will demonstrate the usefulness of both feature importance and RvP analysis for understanding the information content of JWST NIRSpec transmission spectra of warm Neptunes.  The random forest technique has also been applied to ground-based spectra of brown dwarfs at medium spectral resolution \citep{oreshenko20} and ultra-hot Jupiters at high spectral resolution \citep{fisher20}.

\subsection{Motivation IV: planning NIRSpec observations on JWST}

The current study is restricted to transmission spectroscopy at optical to near-infrared wavelengths.  Specifically, we consider the NIRSpec instrument on JWST.\footnote{https://jwst-docs.stsci.edu/near-infrared-spectrograph/nirspec-observing-modes/nirspec-bright-object-time-series-spectroscopy}  In the low-resolution ($\sim 100$) prism mode, NIRSpec has a simultaneous wavelength coverage of 0.6--5.3 $\mu$m.  It is suitable for stars fainter than $J \approx 10$, corresponding to Kepler and fainter TESS targets.  At medium resolution ($\sim 1000$), NIRSpec has four modes: 0.7--1.27 $\mu$m (G140M/F070LP), 0.97--1.84 $\mu$m (G140M/F100LP), 1.66--3.07 $\mu$m (G235M/F170LP) and 2.87--5.10 $\mu$m (G395M/F290LP).  These modes are suitable for stars fainter than $J \approx 6$--8.  Four high-resolution ($\sim 2700$) modes exist as well, but as we will show these do not add much interpretational value, in terms of retrieving elemental and molecular abundances, to what the medium-resolution modes already offer.  Table 1 provides a summary of the JWST NIRSpec modes we will consider in the current study.

%Among the four medium-resolution NIRSpec modes, we wish to determine the one that encode the most information content for hot Neptunes.  Upon identifying this mode, we wish to study if the addition of a second mode adds significant value to the retrieval and to identify the optimal one.

\subsection{Layout of study}

In Section \ref{sect:methods}, we describe our methods of computation.  In Section \ref{sect:results}, we present the results from our information content analyses and also an improved diagnostic for chemical disequilibrium.  In Section \ref{sect:discussion}, we compare our results to those of previous studies and discuss their implications for planning JWST observations.

\section{Methodology}
\label{sect:methods}

\begin{table*}
\label{tab:priors}
\begin{center}
\caption{Retrieved parameters and their prior distributions}
\begin{tabular}{lcccccccc}
\hline
Quantity & Symbol & Units & Range & Prior Type \\
\hline
Temperature & $T$ & K & 800--1200 & uniform \\
Volume mixing ratios & $X_i$ & --- & $10^{-9}$--$10^{-2}$ & log-uniform \\
Cloud extinction coefficient normalisation & $\alpha_0$ & cm$^{-1}$ & $10^{-11}$--$10^{-7}$ & log-uniform \\
Proxy for cloud composition & $Q_0$ & --- & 1--100 & uniform \\
Cloud particle radius & $r_c$ & $\mu$m & $10^{-3}$--$10^3$ & log-uniform \\
\hline
\hline
\end{tabular}\\
\end{center}
\end{table*}

\subsection{Opacities}

\begin{figure*}%[!t]
\begin{center}
\vspace{-0.2in}
\includegraphics[width=1.7\columnwidth]{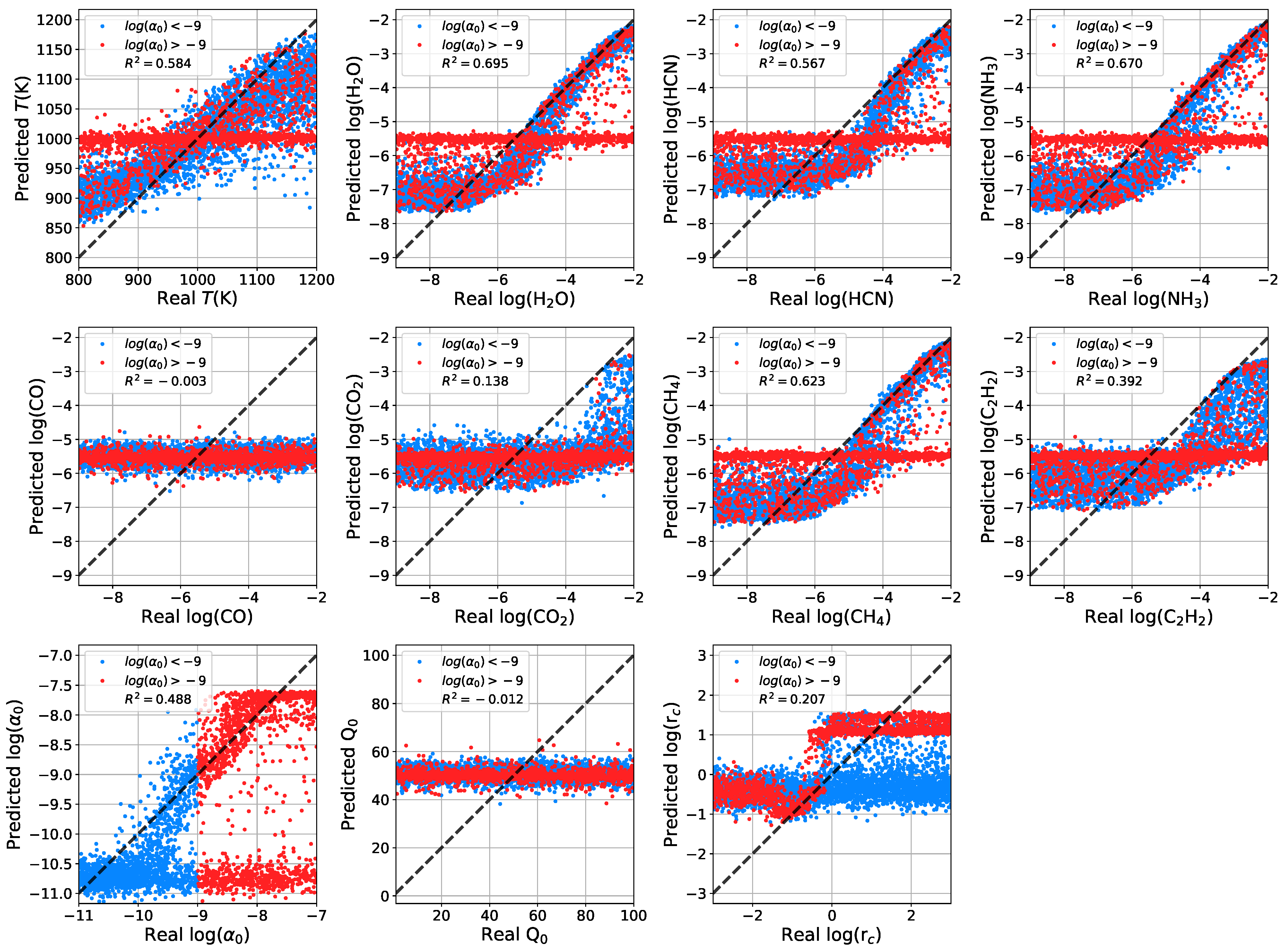}
\includegraphics[width=1.7\columnwidth]{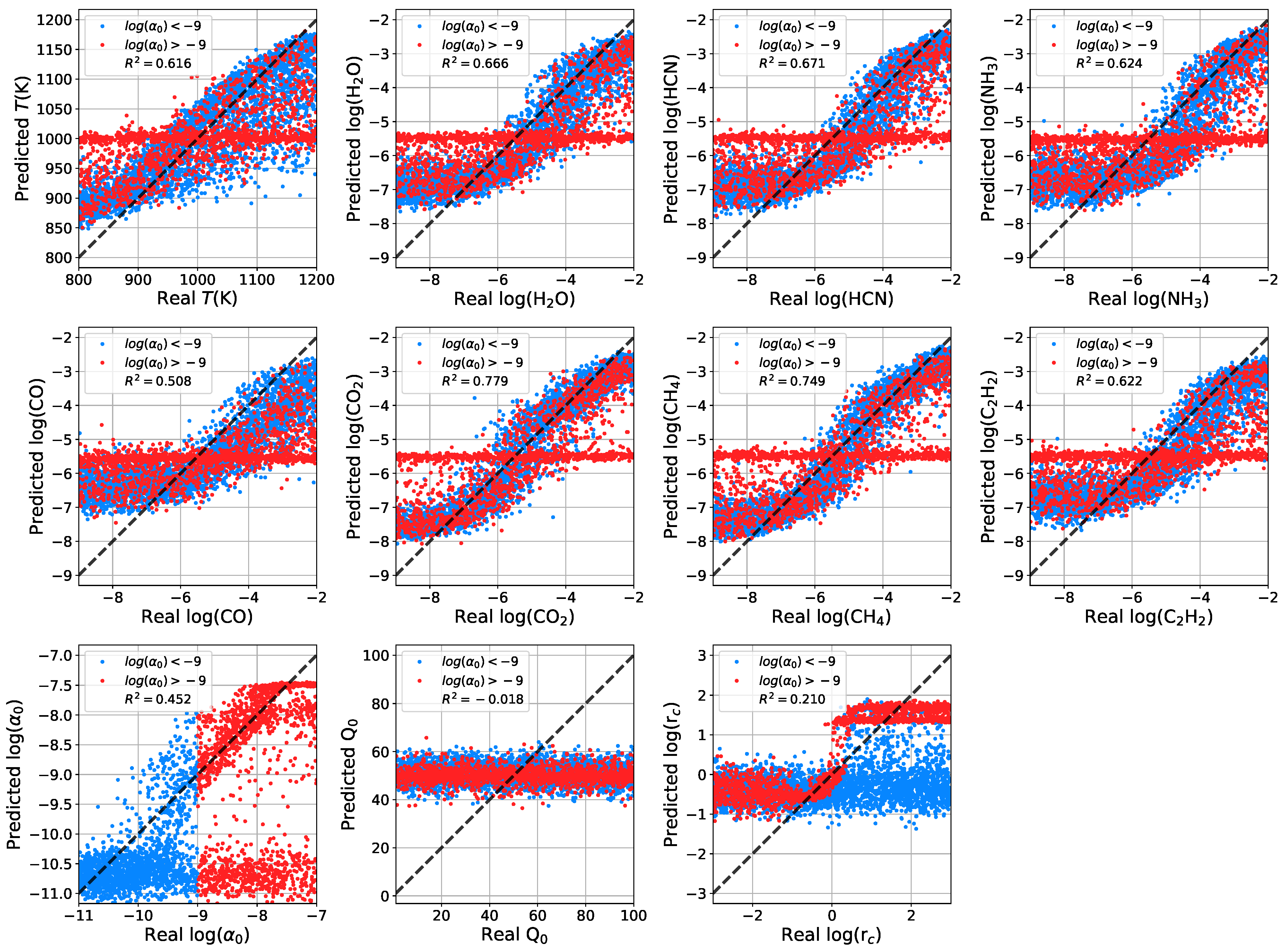}
\end{center}
%\vspace{-0.1in}
\caption{RvP analysis of the medium-resolution M1 (top montage of 11 panels) versus M4 (bottom montage of 11 panels) modes of JWST NIRSpec.  As in Figure \ref{fig:wfc3_rvp}, the blue and red points correspond to cloudfree ($\alpha_0 < 10^{-9}$ cm g$^{-1}$) and cloudy ($\alpha_0 > 10^{-9}$ cm g$^{-1}$) transmission spectra, respectively.  For clarity of presentation, we show only 5000 out of the actual 20,000 mock retrievals performed.}
\vspace{-0.1in}
\label{fig:rvp}
\end{figure*}

\begin{figure*}%[!ht]
\begin{center}
\vspace{-0.2in}
\includegraphics[width=\columnwidth]{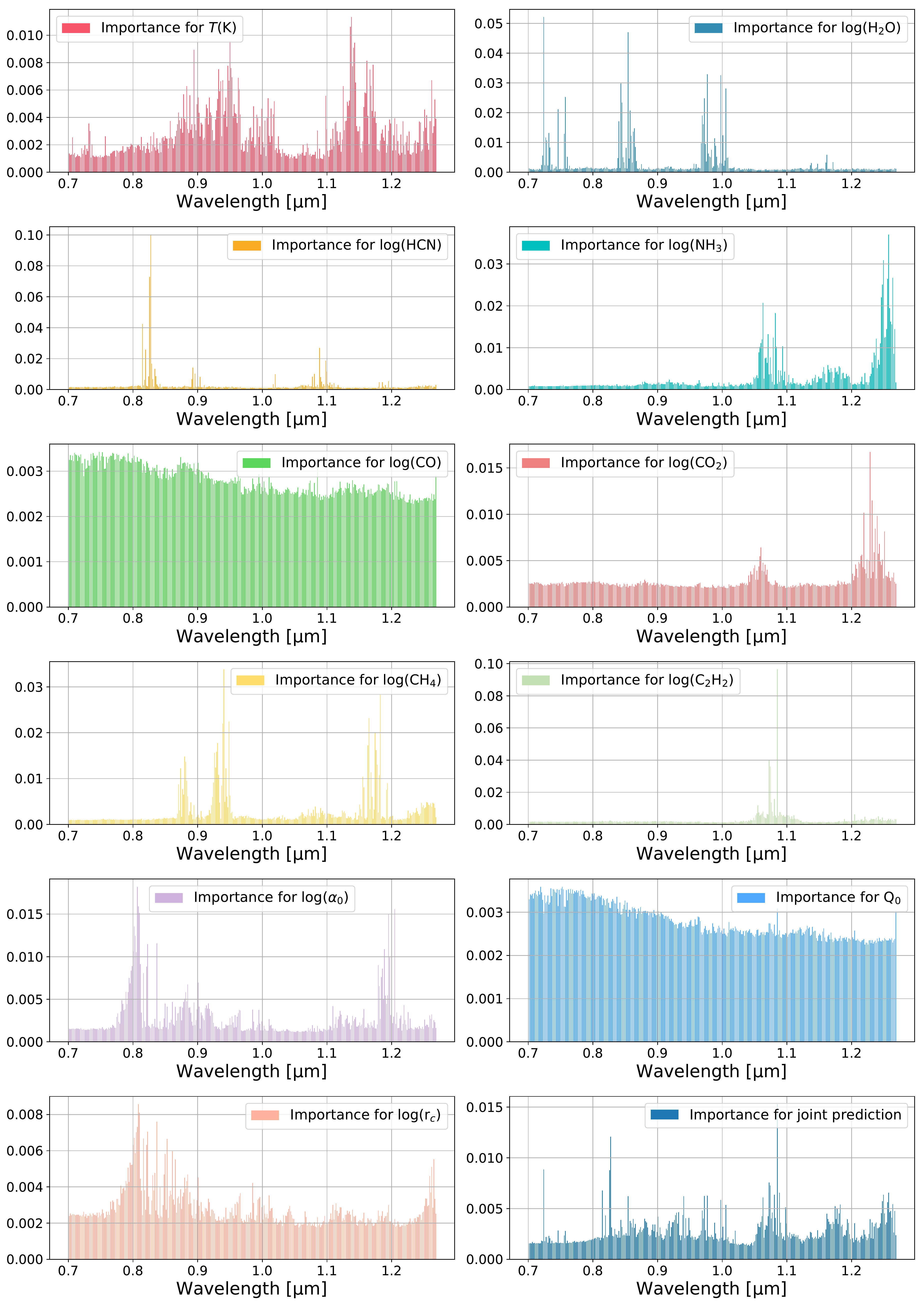}
\includegraphics[width=\columnwidth]{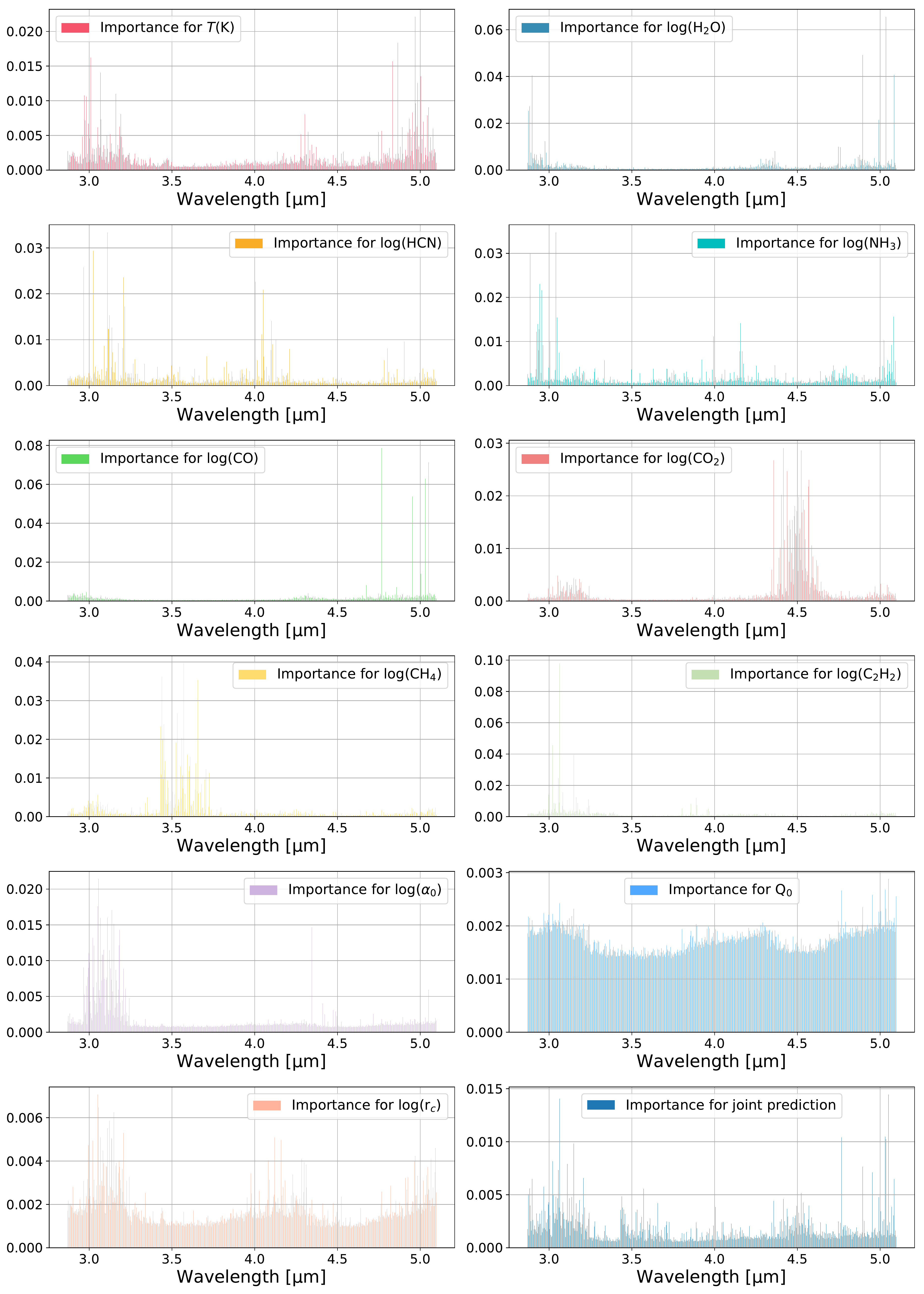}
\end{center}
%\vspace{-0.1in}
\caption{Feature importance analysis of the medium-resolution M1 (left montage of 12 panels) versus M4 (right montage of 12 panels) modes of JWST NIRSpec.}
\vspace{-0.1in}
\label{fig:feature}
\end{figure*}

\subsubsection{Molecules}

The molecular opacities of H$_2$O, HCN, NH$_3$, CO, CO$_2$, CH$_4$ and C$_2$H$_2$ are taken from the \texttt{ExoMol} \citep{barber06,y11,y13,barber14,yt14}, \texttt{HITRAN} \citep{rothman87,rothman92,rothman98,rothman03,rothman05,rothman09,rothman13} and \texttt{HITEMP} \citep{rothman10} spectroscopic databases; the pressure broadening parameters for H$_2$-He mixtures are taken from the \texttt{ExoMol} database.  A review of the spectroscopic databases may be found in \cite{ty17}.  For a review of how to compute opacities given inputs from the spectroscopic databases, we refer the reader to, for example, the appendix of \cite{rothman98}, \cite{gh15}, Chapter 5 of \cite{heng17} and \cite{y18}. All opacities are calculated with the HELIOS-K opacity calculator \citep{gh15} from $10^{-8}$--$10^3$ bar. Table 3 states the details of the opacities, including the range of wavenumbers/wavelengths over which spectroscopic data needed as input exist. When computing transmission spectra, we use opacity sampling at the resolutions stated in Table 1.

\begin{table*}
\label{tab:molecules}
\begin{center}
\caption{Spectroscopic line lists used to generate opacities}
\begin{tabular}{lcccc}
\hline
Molecule & Line list & Shortest wavelength & Wavenumber range & References \\
\hline 
 & & ($\mu$m) & (cm$^{-1}$) & \\
\hline
\hline
H$_2$O & 1H2-16O\_\_POKAZATEL & 0.24 & 0--41200 & \cite{poly18}\\
HCN    & 1H-12C-14N\_\_Harris & 0.57 & 0--17585 & \cite{harris6, barber14} \\ 
NH$_3$ & 14N-1H3\_\_BYTe      & 0.83 & 0--12000 & \cite{y11} \\ 
CO     & 12C-16O\_\_Li2015    & 0.45 & 0--22000 & \cite{li15}\\
CO$_2$ & HITEMP 2010          & 1.04 & 258--9648 & \cite{rothman10}\\ 
CH$_4$ & 12C-1H4\_\_YT10to10  & 0.83 & 0--12100 & \cite{y13,yt14} \\
C$_2$H$_2$ & HITRAN 2016      & 1.01 & 0--9889 & \cite{gordon16}\\
\hline
\hline
\end{tabular}\\
\end{center}
\end{table*}

\subsubsection{Clouds}

In the current study, we will use the terms ``cloud", ``haze" and ``aerosol" interchangeably, based on the reasoning that while these terms may reflect different formation pathways, the effects on a spectrum follow a common phenomenological treatment.  There is no consensus on the use of these terms: Earth scientists use ``haze" versus ``cloud" as a measure of particle size, while planetary scientists use these terms to refer to photochemical and thermochemical formation origins, respectively.

If a cloud consists of spherical particles of a single radius (i.e., a monodisperse cloud), then its cross section is
\begin{equation}
\sigma_c = Q \pi r_c^2,
\end{equation}
where $Q$ is the extinction efficiency.  It may be computed using Mie theory \citep{mie}.  \cite{kh18} use the open-source \texttt{LX\_MIE} Mie code to calibrate a convenient fitting function for $Q$, 
\begin{equation}
Q = \frac{Q_1}{Q_0 x^{-a} + x^{0.2}},
\end{equation}
where $Q_1 \approx 4$ \citep{kh18}, the dimensionless size parameter is $x = 2 \pi r_c / \lambda$ and $\lambda$ is the wavelength.  This fitting function smoothly transitions between the regimes of small ($x \ll 1$; Rayleigh and non-gray continuum) and large ($x \gg 1$, gray continuum) particles.  For simplicity, we assume $a=4$; see Table 2 of \cite{kh18} for the values of $a$ as a function of the composition.  Refractory and volatile condensates correspond to $Q_0 \sim 10$ and $\sim 100$, respectively (see Table 2 of \citealt{kh18}).  The cloud extinction coefficient is assumed to be uniform along the transit chord.

It is worth noting that this simplified treatment of the cloud cross section does not capture composition-specific spectral features (e.g., \citealt{cushing09,lee14}).

As it is calibrated on first-principles calculations, our treatment of clouds is an improvement over the gray cloud assumption of \cite{howe17} and the approach of \cite{greene16} and \cite{bl17}, who used a combination of a ``cloud top pressure" (for gray clouds) and a power-law parametrisation (for non-gray clouds).

\subsubsection{Total extinction coefficient}
The total extinction coefficient is
\begin{equation}
\alpha = \alpha_c + \sum_i \frac{m_i}{m} X_i \kappa_i \rho,
\end{equation}
where $m_i$ is the mass, $X_i$ is the volume mixing ratio, $\kappa_i$ is the opacity of each molecule. The sum is performed over all of the molecules in the system.  The mass density and mean molecular mass of the atmosphere are given by $\rho$ and $m$, respectively.  The extinction coefficient associated with clouds is written as
\begin{equation}
\alpha_c = \frac{\alpha_0}{Q_0 x^{-4} + x^{0.2}}.
\end{equation}
The mean molecular mass, cloud volume mixing ratio ($X_c$) and $Q_1$ are subsumed into a single fitting parameter,
\begin{equation}
\alpha_0 \propto \frac{Q_1 \pi r_c^2 X_c}{m}.
\end{equation}

\subsection{Transmission spectra}
\label{sec:transmission_spectra}

Consistent with \cite{greene16} and \cite{howe17}, we assume isothermal, non-isobaric transit chords.\footnote{It is not equivalent to assuming an isothermal atmosphere.  Rather, it is the assumption that the region of the atmosphere probed by transmission spectroscopy is isothermal over the wavelength (and hence pressure) range considered.}  We use the \texttt{HELIOS-O} code to compute transmission spectra \citep{gaidos17,bower19}.  Each model atmosphere is divided into 150 annuli in pressure ($P$) from $10^{-8}$--$10$ bar.  The limit of 10 bar is chosen to ensure that the atmosphere is fully opaque at the lower boundary and has no bearing on the final outcome of the calculation.

At each wavelength, the slant optical depth is computed using \citep{brown01}
\begin{equation}
\tau = \int^{\infty}_{-\infty} \alpha ~dx,
\end{equation}
where $x$ is the spatial coordinate along the line of sight.  The transmission function along each line of sight is 
\begin{equation}
{\cal T} = e^{-\tau}.
\end{equation}
Integrating along the radial coordinate yields the transit depth \citep{brown01},
\begin{equation}
\left( \frac{R}{R_\star} \right)^2 = \frac{1}{R^2_\star} \int^{\infty}_0 ~2\pi r \left( 1 - {\cal T} \right) ~dr.
\end{equation}

JWST spectra are expected to encode enough information \citep{fh18} to break the normalisation degeneracy \citep{bs12,g14,barstow15,hk17,heng19}.  Nevertheless, we account for this degeneracy by matching the computed white-light radius of each model to the measured one ($R=0.3767 ~R_{\rm J}$ from 0.5--1.0 $\mu$m; \citealt{torres08}).

%we do not retrieve for either the reference pressure ($P_0$) or the reference transit radius ($R_0$) and instead fix their values to $R_0 = 0.3532 ~R_{\rm J}$ (Table 3 of \citealt{fh18}) and $P_0=10$ bar.  

Across the wavenumber range of $1/\lambda=1800$--17000 cm$^{-1}$ (wavelength range of 0.6--5.5 $\mu$m), we assume a uniform spacing in $\log{(1/\lambda)}$ corresponding to 6700 points, such that the spectral resolution is approximately constant with an average value of 3000. The spectra are then restricted in wavelength and binned down to a spectral resolution of 100, 600 or 1000, depending on which modes of NIRSpec one is studying (see Table 1).  An optimistic, photon-limited uncertainty of 20 ppm per data point is assumed, consistent with \cite{greene16}.  The intention is to identify the possible weaknesses of each NIRSpec mode even under idealized conditions.

On a desktop computer (Intel Core i9-7960X CPU), it takes \texttt{HELIOS-O}, which is written in the C++ programming language, about 1 second to compute each model. For the entire grid of 100,000 models, this amounts to about 30 hours of computing time.

\subsection{Random forest retrieval}
\label{subsect:rf}

The ``random forest" is a classical, supervised method of machine learning \citep{ho98,breiman01}.  It belongs to a class of methods known as Approximate Bayesian Computation (ABC).  Within the ABC framework, it has been demonstrated that one may compute approximate posterior distributions and perform model comparison via computation of the Bayesian evidence \citep{sisson19}.

As is appropriate for continuous quantities such as transit depths or radii, a regression tree (rather than a decision tree) is used to classify transmission spectra with different sets of parameter values (treated as ``labels"; \citealt{mn18}).  A bootstrapping method is used to generate an uncorrelated forest of regression trees, and the combined output of the random forest yields the posterior distributions of parameters \citep{crimi11}.  Following \cite{fisher20}, we take as output all of the entries in a leaf, rather than the average of the leaf, as the sampled posterior distribution of a parameter.

As demonstrated by \cite{mn18}, the random forest produces two additional diagnostics: feature importance plots, which quantify the relative importance of each data point in the transmission spectrum for constraining each parameter; and RvP plots, which quantify the degree to which each parameter may be predicted in mock retrievals given the noise model.  The RvP analysis is essentially an efficient way to generate large suites ($\sim 10^4$) of mock retrievals, which is computationally challenging to accomplish using standard retrieval methods (e.g., \citealt{barstow15}).

The range of values of the model parameters, as well as the assumptions on their prior distributions, are stated in Table \ref{tab:priors}. Each parameter is randomly drawn from its prior and a noise-free transmission spectrum is generated, as explained in Section \ref{sec:transmission_spectra}. In order to add noise, each point in the synthetic spectrum is assumed to follow a Gaussian distribution with a standard deviation of 20 ppm. The points are then randomly sampled from these distributions, centred on their noise-free values. 

This is repeated to build a grid of 100,000 models for the forest, split into 80,000 for training and 20,000 for testing. The random forest consists of 1000 trees. Tree splitting is performed using the following steps: the range of values of each parameter is normalized such that its maximum value is 100; tree splitting ceases when the change in total variance of the parameter values (as a node is split into two branches) is less than a stated tolerance, which is set to 0.01.  Each time a tree is split, a random subset of $\sim\sqrt{N}$ points is used, where $N$ is the total number of spectral points, to reduce biases. Tree pruning methods are not used. The implementations of the random forest method and $R^2$ metric are from the open-source \texttt{scikit.learn} library \citep{pedregosa11} in the \texttt{Python} programming language.

On a desktop computer (Intel Core i7 CPU), it takes \texttt{HELA}, which is written in the \texttt{Python} programming language, about 10 minutes to train the random forest.

\section{Results}
\label{sect:results}

As an illustration, we will use the example of GJ 436b for our calculations: the GJ 436 star has a stellar radius of $R_\star = 0.455 ~R_\odot$ and GJ 436b has a surface  of $g=1318$ cm s$^{-2}$ \citep{vb12}.  The qualitative conclusions of our study do not depend on the choice of these parameter values.

\begin{figure*}
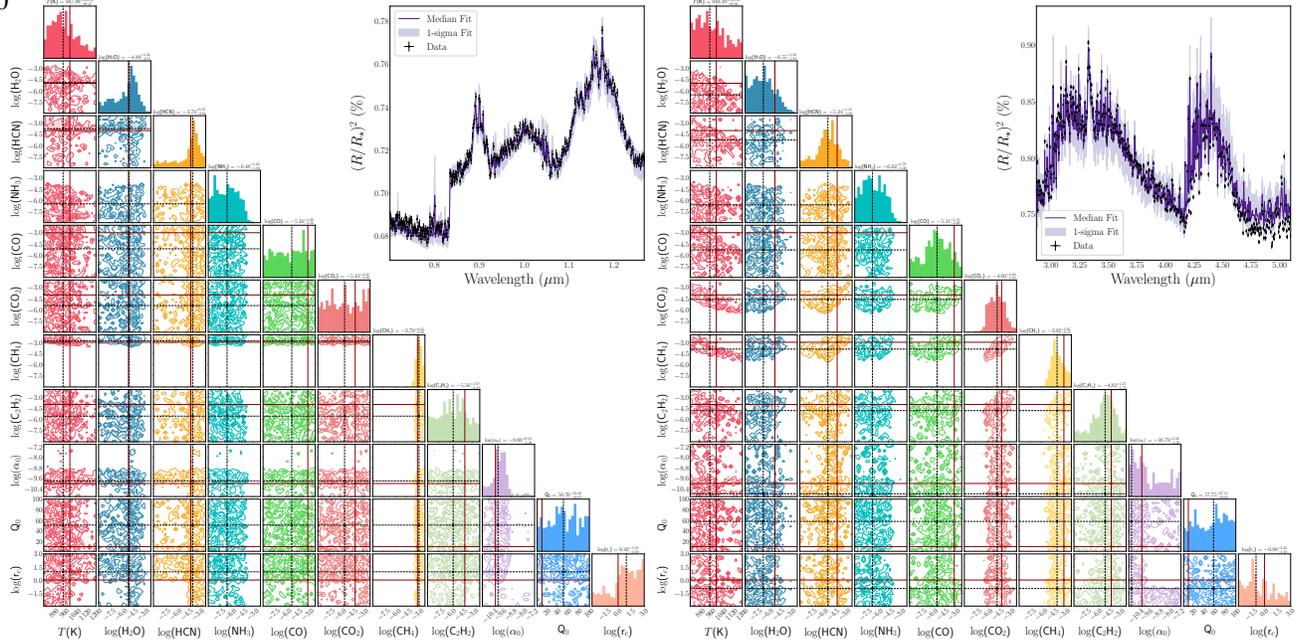
%[!ht]
\begin{center}
\vspace{-0.2in}
\includegraphics[width=\columnwidth]{f5a.pdf}
\includegraphics[width=\columnwidth]{f5b.pdf}
\end{center}
%\vspace{-0.1in}
\caption{Posterior distributions from mock retrievals of a carbon-rich (water-poor) atmosphere with $T=1000$ K, $X_{\rm CO}=X_{\rm CH_4}=10^{-3}$, $X_{\rm HCN}=X_{\rm C_2H_2}=X_{\rm CO_2}=10^{-4}$, $X_{\rm H_2O}=10^{-5}$, $\alpha_0=10^{-10}$ cm$^{-1}$, $Q_0=10$ and $r_c=1$ $\mu$m.  The left and right montages are for the M1 and M4 modes, respectively. The kink in the synthetic spectrum associated with the M1 mode is due to the non-existence of CH$_4$ line list data at bluer wavelengths.  The vertical black dotted lines show the median value of each posterior distribution.  Wherever applicable, the vertical red solid lines show the truth values of a parameter.}
\vspace{-0.1in}
\label{fig:carbon}
\end{figure*}

\begin{figure*}%[!ht]
\begin{center}
%\vspace{-0.2in}
\includegraphics[width=\columnwidth]{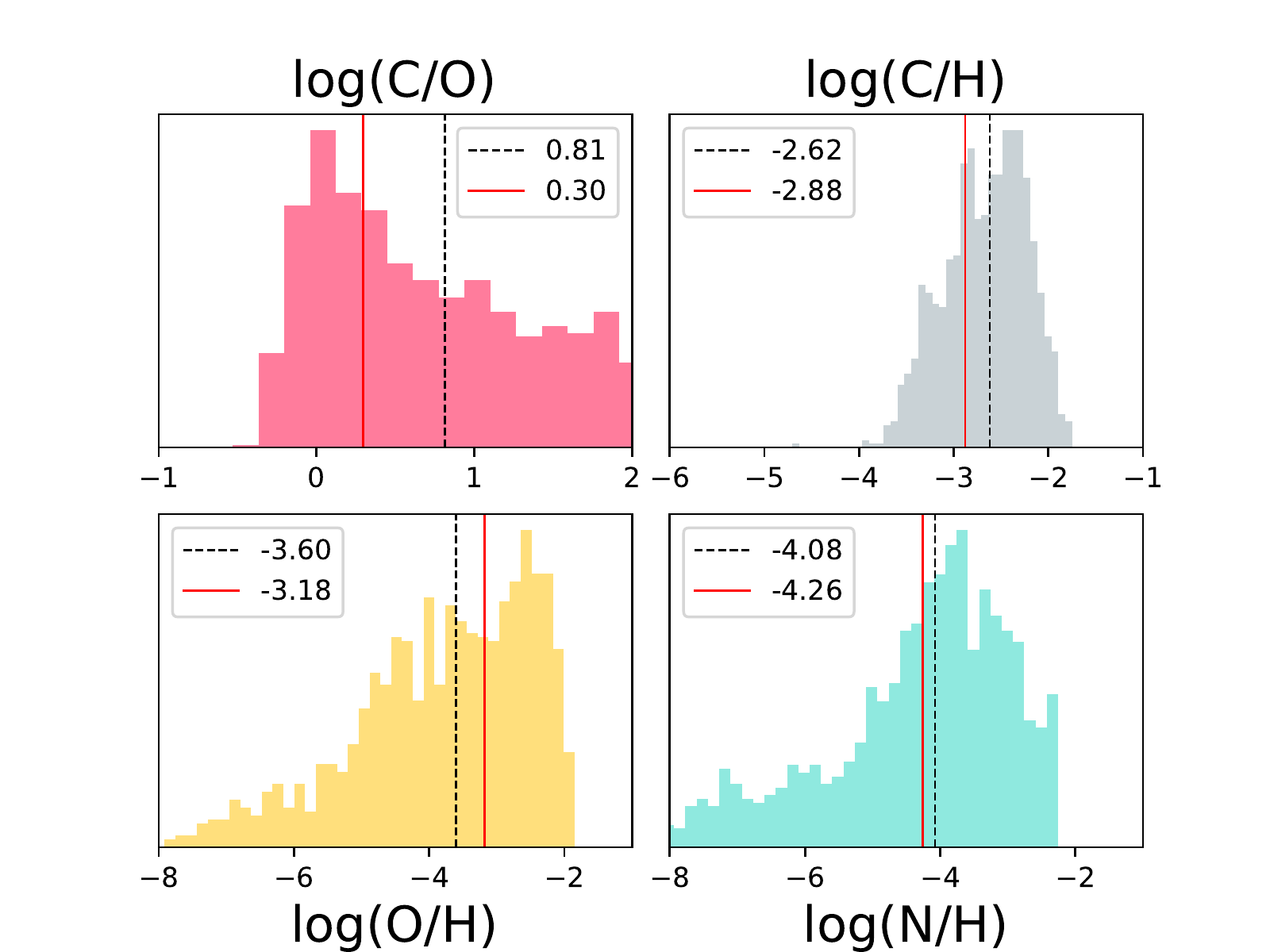}
\includegraphics[width=\columnwidth]{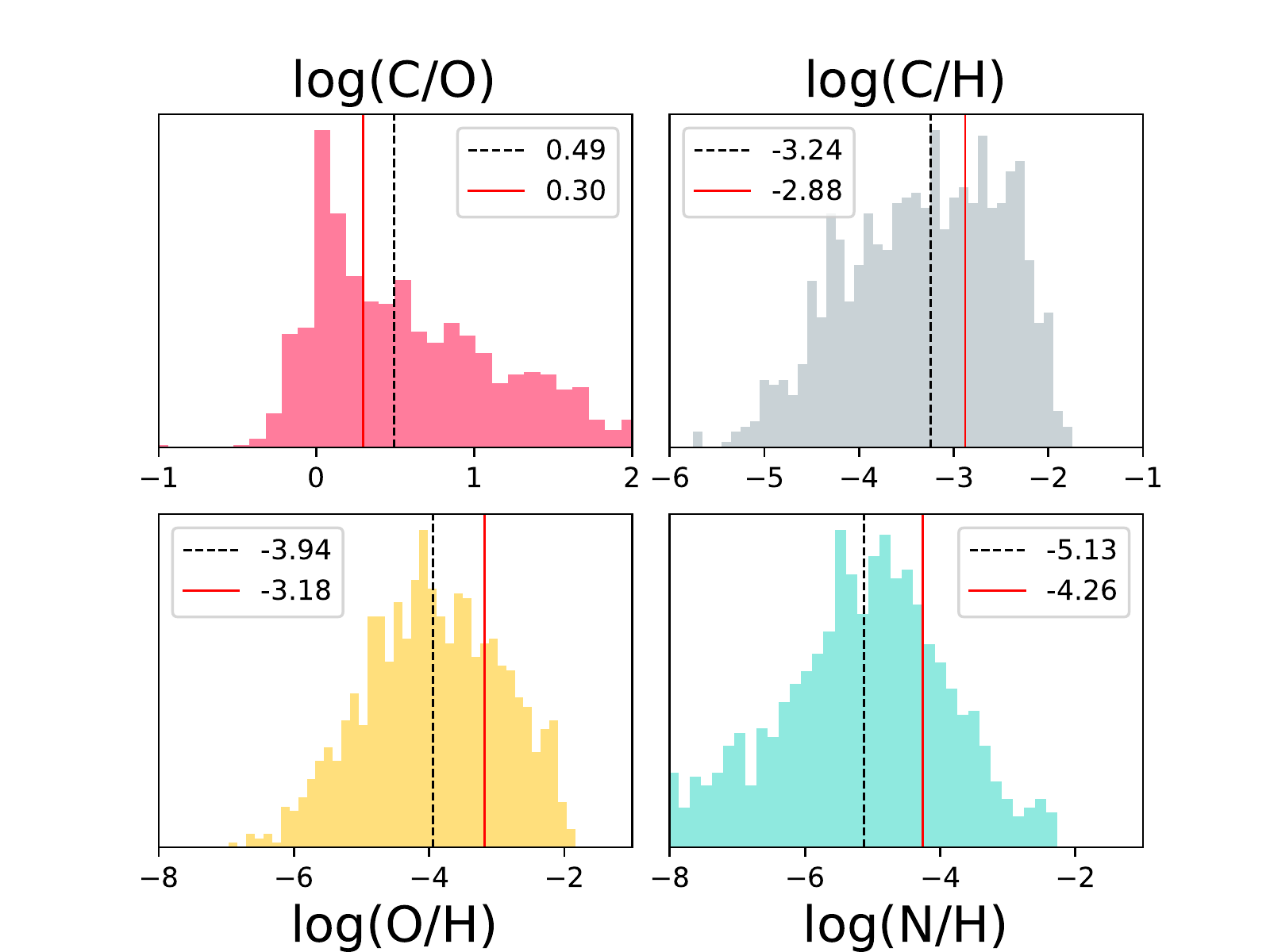}
\end{center}
%\vspace{-0.1in}
\caption{Posterior distributions of C/O, C/H, O/H and N/H, computed by post-processing the primary posteriors obtained in Figure \ref{fig:carbon}.  The left and right montages are for the M1 and M4 modes, respectively.  The vertical black dotted lines show the median value of each posterior distribution, which are also indicated numerically in each panel (as the logarithm of each quantity).  The vertical red solid lines correspond to the truth values.}
\vspace{-0.1in}
\label{fig:carbon2}
\end{figure*}

\subsection{RvP analysis of different JWST NIRSpec modes}
\label{subsect:predictability}

Table 1 lists the 4 medium-resolution modes of JWST NIRSpec. The expectation is that the M1 (0.7--1.27 $\mu$m) and M2 (0.97--1.84 $\mu$m) modes, which probe a collective wavelength range similar to the WFC3 instrument of HST, encode the most information on cloud properties (e.g., \citealt{lec08}), but may be insensitive to important carbon-bearing molecules such as CO and CO$_2$. Therefore, we begin the discussion by comparing the M1 and M4 (2.87--5.10 $\mu$m) modes.  

Figure \ref{fig:rvp} shows the outcomes of performing 20,000 mock retrievals for each of the modes in turn.  For clarity of presentation (and with no loss of generality), we display only 5000 out of the 20,000 mock retrievals.  It is emphasized again that the random forest reports the \textit{mean} predicted value of each parameter.\footnote{The median value may also be reported, which is the approach followed by \cite{fisher20}.}  Based on the similar ${\cal R}^2$ values obtained, the M1 and M4 modes do comparably well at constraining the H$_2$O and NH$_3$ abundances, as well as the temperature. The M4 mode outperforms the M1 mode by more than 0.1 in ${\cal R}^2$ value for constraining the abundances of HCN, CH$_4$ and C$_2$H$_2$. As demonstrated by the low ${\cal R}^2$ values, the M1 mode is insensitive to CO$_2$ (${\cal R}^2=0.138$) and essentially blind to CO (${\cal R}^2=-0.003$). The M4 mode offers drastic improvements on constraining CO (${\cal R}^2=0.508$) and CO$_2$ (${\cal R}^2=0.779$) due to their spectral features across 4--5 $\mu$m (Figure \ref{fig:feature}).

When a mock retrieval fails to predict the value of a given parameter, the RvP analysis returns values that are the mean of the range considered.  In the case of CO, since the range of volume mixing ratios considered is $10^{-9}$--$10^{-2}$ (in log-uniform spacing), the random forest returns $X_{\rm CO}=10^{-6}$--$10^{-5}$ for the M1 mode.  In other RvP plots where the predicted values of the parameters level off at a value that is below the mean of the range considered, these indicate the minimum or threshold value of a parameter that can be constrained given the noise model. For example, $X_{\rm H_2O} \gtrsim 10^{-7}$ for both the M1 and M4 modes.  Generally, volume mixing ratios as low as $\sim 10^{-8}$ may be constrained given the assumed 20 ppm noise floor.

In Figure \ref{fig:rvp}, the points have been color-coded blue ($\alpha_0 < 10^{-9}$ cm$^{-1}$) or red ($\alpha_0 > 10^{-9}$ cm$^{-1}$) to correspond to cloudfree or cloudy atmospheres, respectively.  This threshold value of $\alpha_0$ was obtained by trial-and-error and is guided mainly by inspecting the RvP behavior of both $\alpha_0$ and $r_c$. The bimodal behavior of $\alpha_0$ above this threshold is an indication of the degeneracy between the degree of cloudiness and the molecular abundances.  The trend of $r_c$ leveling off at $\gtrsim 1$ $\mu$m is the outcome of the cloud opacity becoming gray/constant as the cloud particles become large compared to the wavelengths probed.  This trend is consistent with the basic principles of Mie theory.  In all of the RvP plots of the molecular abundances and temperature, a subpopulation of the red (cloudy) points cluster in the middle of the range of values considered, indicating that the random forest does not predict a value for the given parameter.

The M1 and M4 modes constrain $\alpha_0$ (${\cal R}^2=0.488$ versus $0.452$) and $r_c$ (${\cal R}^2=0.207$ versus $0.210$) almost equally well.  Both the M1 and M4 modes have no sensitivity to the cloud composition (via $Q_0$; ${\cal R}^2 \approx 0$), which implies that it is challenging to identify cloud composition by constraining changes in the gradient of the spectral continuum alone.  It does not rule out the possibility that higher-order spectral features that are composition-specific may retain constraining power (e.g., \citealt{cushing09,lee14}).

For completeness, the RvP plots of the M2, M3 and L modes are included in the Appendix as Figures \ref{fig:M2_RvP}, \ref{fig:M3_RvP} and \ref{fig:L_RvP}, respectively.  The M2 mode exhibits similar behavior as the M1 mode in that it is somewhat insensitive to CO$_2$ (${\cal R}^2=0.322$) and nearly blind to CO (${\cal R}^2=0.039$).  The M3 mode (1.66--3.07 $\mu$m) is blind to CO (${\cal R}^2=0.075$), but sensitive to CO$_2$ (${\cal R}^2=0.669$).  The L mode has good sensitivity to CO$_2$ (${\cal R}^2=0.763$), but is largely insensitive to CO (${\cal R}^2=0.171$).  Section \ref{subsect:compare} and Figure \ref{fig:compare} performs a detailed comparison of the ${\cal R}^2$ values of every parameter for all of the modes considered in the present study.

\subsection{Feature importance analysis of JWST NIRSpec modes}

Figure \ref{fig:feature} shows the feature importance analysis of the M1 versus M4 modes.  Each panel shows the fractional importance of each data point for constraining a given parameter.  It cannot be over-emphasized that the feature importance values cannot be compared between panels, because the entries are normalised such that they add up to unity \textit{within the same panel}.

The feature importance analysis of the M4 mode reproduces our intuition about the warm \textit{Spitzer Space Telescope} channels.  Channel 1 of the IRAC instrument, which ranges from about 3.1--3.9 $\mu$m and is often quoted as the ``3.6 $\mu$m channel", probes several spectral features of methane (e.g., \citealt{sudarsky03,fortney05,fortney06,fortney10}).  Channel 2 of IRAC, which ranges from about 3.9--5.1 $\mu$m and is often quoted as the ``4.5 $\mu$m channel", probes carbon monoxide (e.g., \citealt{sudarsky03,fortney05,fortney06,fortney10,char08}).  It is consistent with the narrative that the flux ratios of these channels probe the relative abundances of CH$_4$ to CO, and is thus a measure of disequilibrium chemistry (e.g., \citealt{ms11,moses11}).

Other properties are less apparent without detailed scrutiny of the feature importance plots.  Generally, molecules such as H$_2$O, HCN, NH$_3$, CH$_4$ and C$_2$H$_2$ have multiple spectral lines distributed across the wavelength ranges of both the M1 and M4 modes.  For the M4 mode, there are strong CO$_2$ features between 4--5 $\mu$m.  It also encompasses a CO feature at 4.7 $\mu$m, which explains the ability of the 4.5 $\mu$m channel of IRAC to constrain carbon monoxide.  The M1 and M4 modes are equally good at constraining $\alpha_0$ and the cloud particle radius (based on comparing the ${\cal R}^2$ values as discussed earlier), but these constraints come from different wavelength regions.  

Parameters associated with ${\cal R}^2 \approx 0$ typically have almost equal feature importance distributed across wavelength, as is the case for $Q_0$ (both M1 and M4) and CO (only M1).  For completeness, we include in the Appendix the feature importance plots of the M2, M3 and L modes in Figures \ref{fig:M2_feature}, \ref{fig:M3_feature} and \ref{fig:L_feature}, respectively.

\subsection{Posterior distributions from mock retrievals}
\label{subsect:posteriors}

As an illustration, we consider a case study that is motivated by qualitative trends in gaseous, equilibrium chemistry at $\sim 1000$ K (e.g., \citealt{moses11,moses13,madhu12,ht16}): A carbon-rich (water-poor) atmosphere consisting of $X_{\rm CO}=X_{\rm CH_4}=10^{-3}$, $X_{\rm HCN}=X_{\rm C_2H_2}=X_{\rm CO_2}=10^{-4}$ and $X_{\rm H_2O}=10^{-5}$, which corresponds to $\mbox{C/O} \approx 1.98$ or $\log{\mbox{C/O}} \approx 0.30$.  For illustration, we assume $T=1000$ K, $\alpha_0=10^{-10}$ cm$^{-1}$, $Q_0=10$ and $r_c=1$ $\mu$m.

Consistent with the insensitivity of the M1 mode to CO, CO$_2$ and C$_2$H$_2$, the posterior distributions of these molecules are unconstrained (Figure \ref{fig:carbon}).  The M4 mode does surprisingly poorly on CO, but this is because its spectral lines are being masked by those of CO$_2$ and CH$_4$ (see Appendix).  Both modes obtain only an upper limit for NH$_3$, which is absent from this model atmosphere.  Overall, the M4 mode does somewhat better at retrieving the C/O ratio compared to the M1 mode (Figure \ref{fig:carbon2}).

%For the carbon-rich atmosphere (Figure \ref{fig:carbon}), the M4 mode is now superior to the M3 mode, because of its enhanced ability to constrain the abundances of CO, CO$_2$ and CH$_4$.  It results in a more accurate estimate of C/O using the M4 mode.

Identifying the minimum set of molecules needed to explain a spectrum may be achieved using Bayesian model comparison (e.g., \citealt{bs12,waldmann15,fh18}) or deep learning methods (e.g., \citealt{waldmann16}), which are beyond the scope of the present study.

\subsection{An alternative diagnostic for detecting chemical disequilibrium}

\begin{figure}%[!t]
\begin{center}
%\vspace{0.1in}
\includegraphics[width=\columnwidth]{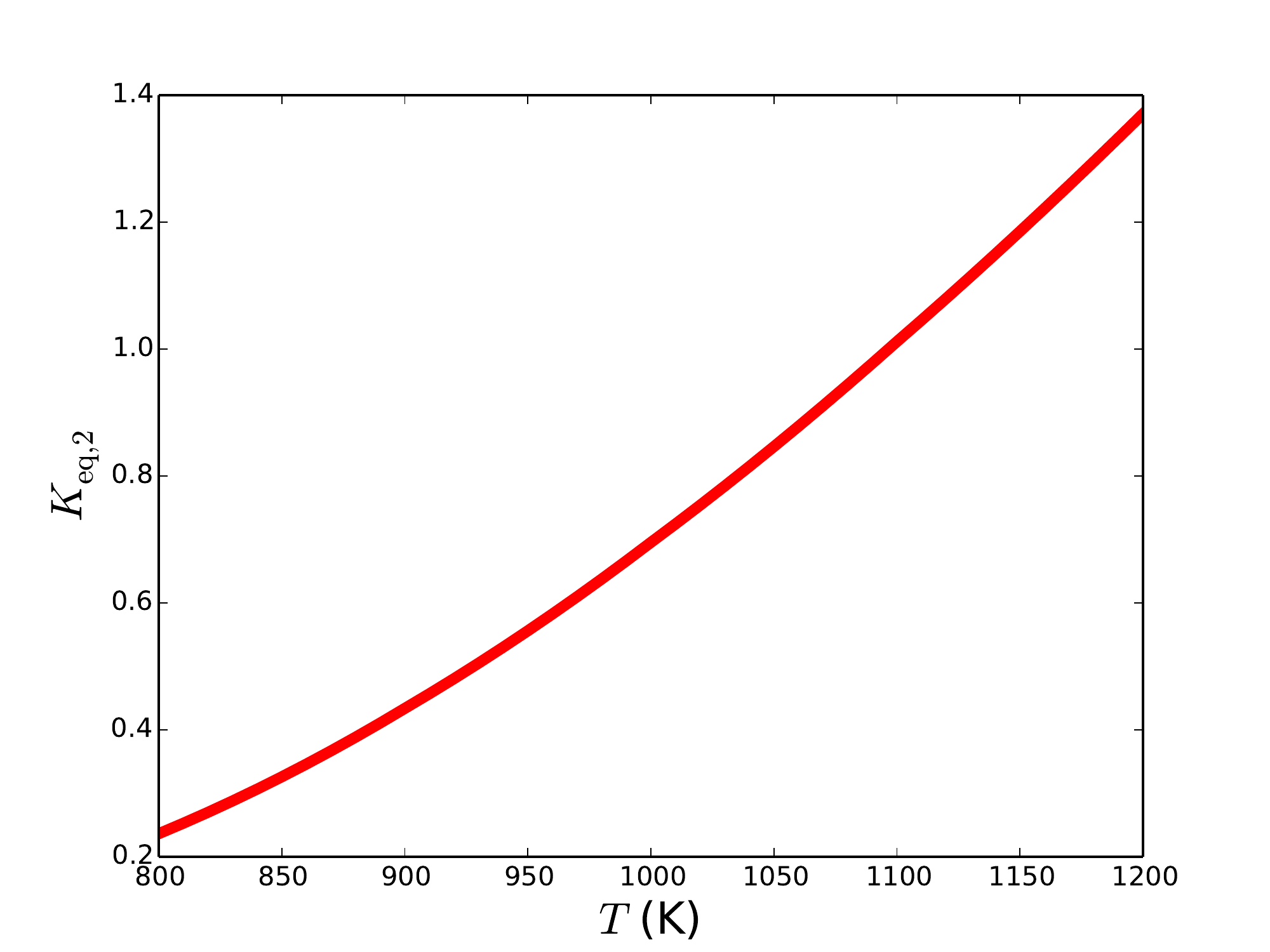}
\end{center}
%\vspace{-0.1in}
\caption{Equilibrium constant associated with the chemical reaction $\mbox{CO}_2 + \mbox{H}_2 \leftrightarrows \mbox{CO} + \mbox{H}_2\mbox{O}$. There is no dependence on pressure.}
%\vspace{-0.1in}
\label{fig:keq2}
\end{figure}

\begin{figure}%[!t]
\begin{center}
%\vspace{0.1in}
\includegraphics[width=\columnwidth]{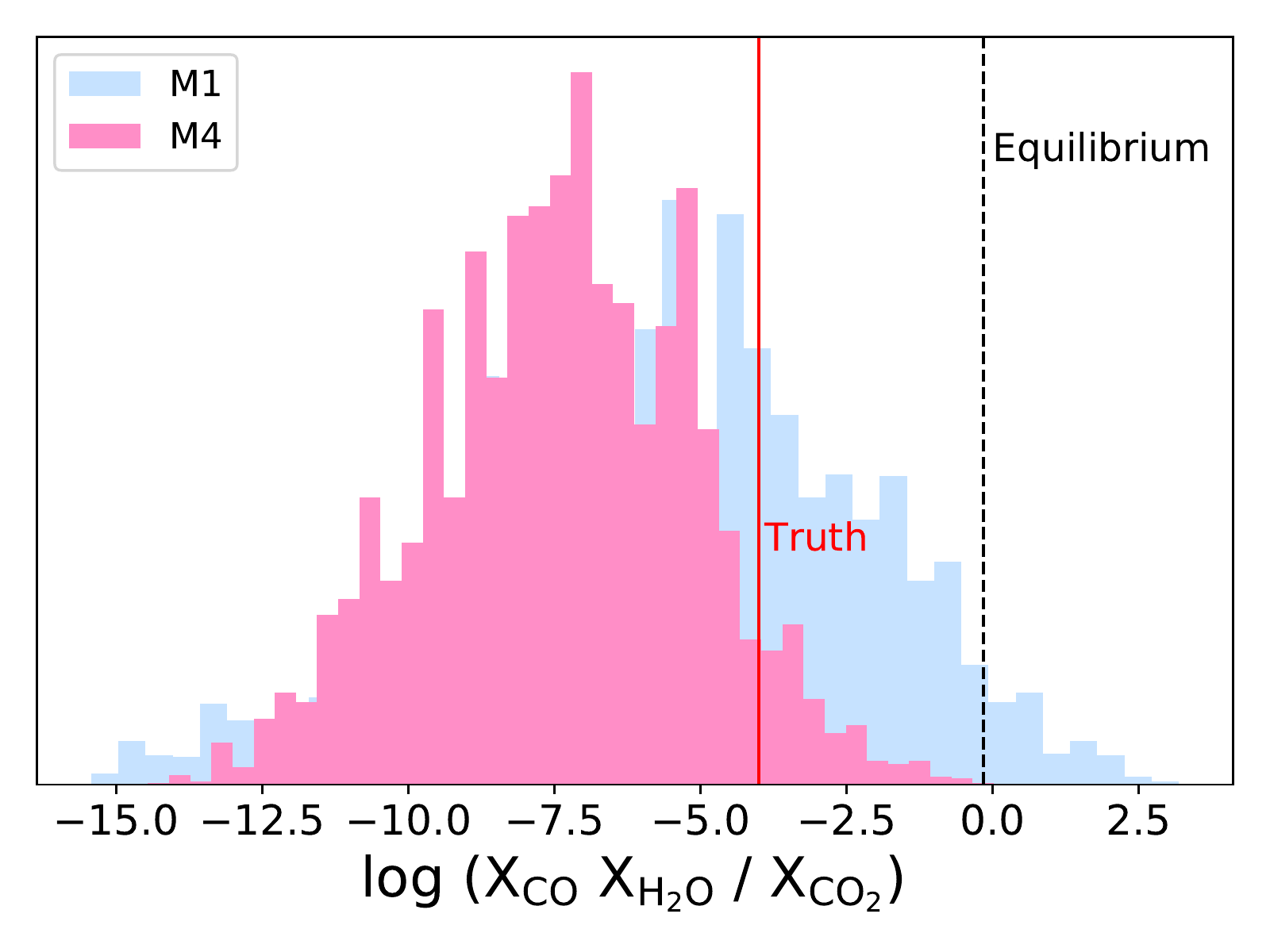}
\end{center}
%\vspace{-0.1in}
\caption{Posterior distributions of the chemical disequilibrium diagnostic corresponding to the carbon-rich case study presented in Figure \ref{fig:carbon}.  The solid vertical line is the truth value, while the dashed vertical line is the value in chemical equilibrium (0.695).}
%\vspace{-0.1in}
\label{fig:diagnostic}
\end{figure}

\cite{ly13} previously proposed a simple diagnostic for identifying chemical disequilibrium in an atmosphere, based on measuring the volume mixing ratios associated with the following chemical reaction (e.g., \citealt{moses11}),
\begin{equation}
\mbox{CH}_4 + \mbox{H}_2\mbox{O} \leftrightarrows \mbox{CO} + 3 \mbox{H}_2.
\end{equation}
When rewritten in the formalism of \cite{ht16}, equation (2) of \cite{ly13} is the reciprocal of
\begin{equation}
\frac{X_{\rm CO}}{X_{\rm CH_4} X_{\rm H_2O}} \left( \frac{P}{P_0} \right)^2,
\label{eq:ly13}
\end{equation}
where $P_0=1$ bar is an arbitrary reference pressure.  If the transit chord probed is in chemical equilibrium, then the preceding expression \textit{is} the equilibrium constant,
\begin{equation}
K_{\rm eq} = \exp{\left(-\frac{\Delta \tilde{G}_{0,1}}{{\cal R}_{\rm univ} T} \right)},
\end{equation}
where ${\cal R}_{\rm univ} = 8.3144621$ J K$^{-1}$ mol$^{-1}$ is the universal gas constant and $\Delta \tilde{G}_{0,1}$ is the molar Gibbs free energy of the reaction (at the reference pressure) tabulated in the \texttt{JANAF} database\footnote{\texttt{https://janaf.nist.gov}} and listed in, for example, Table 2 of \cite{hl16}.  

The key idea proposed by \cite{ly13} is to retrieve for the volume mixing ratios ($X_{\rm CO}, X_{\rm CH_4}, X_{\rm H_2O}$) and obtain an estimate for equation (\ref{eq:ly13}).  If this estimate disagrees with $K_{\rm eq}$ (which requires the retrieved temperature as an input), then the region of the atmosphere probed by transmission spectroscopy is in chemical disequilibrium.  The major uncertainty with this approach is that the pressure probed in transmission ($P$) needs to be accurately and precisely known, especially since it appears as the square of itself in equation (\ref{eq:ly13}).  See Section 6.3 of \cite{greene16} for a critique of \cite{ly13}.

Using the same concept, we propose to focus on another chemical reaction (e.g., \citealt{moses11}), 
\begin{equation}
\mbox{CO}_2 + \mbox{H}_2 \leftrightarrows \mbox{CO} + \mbox{H}_2\mbox{O},
\end{equation}
where the corresponding combination of volume mixing ratios has no dependence on pressure (e.g., \citealt{ht16}),
\begin{equation}
\frac{X_{\rm CO} X_{\rm H_2O}}{X_{\rm CO_2}},
\label{eq:diag1}
\end{equation}
because the number of molecules associated with the reactants and products is the same.  As we will see in Section \ref{subsect:compare}, only the M4 mode of JWST NIRSpec is highly sensitive to the presence of CO, CO$_2$ and H$_2$O.  By retrieving for their mixing ratios and obtaining an estimate for the preceding expression, one may then compare it to the corresponding equilibrium constant,
\begin{equation}
K_{\rm eq,2} = \exp{\left(-\frac{\Delta \tilde{G}_{0,2}}{{\cal R}_{\rm univ} T} \right)},
\label{eq:diag2}
\end{equation}
where $\Delta \tilde{G}_{0,2}$ is again listed in Table 2 of \cite{hl16}.  In chemical equilibrium, equations (\ref{eq:diag1}) and (\ref{eq:diag2}) are equal.  Figure \ref{fig:keq2} shows that $K_{\rm eq,2}$ varies by a factor of about 7 from 800--1200 K.  

To accurately employ this diagnostic, the spectra measured using JWST NIRSpec would have to be of a good enough quality to demonstrate that $X_{\rm CO} X_{\rm H_2O}/X_{\rm CO_2}$ is sufficiently different from $K_{\rm eq,2}$.  In Figure \ref{fig:diagnostic}, we show as an illustration the pair of posterior distributions of $X_{\rm CO} X_{\rm H_2O}/X_{\rm CO_2}$ from retrievals on a mock spectrum corresponding to the M1 and M4 modes for the carbon-rich case study considered in Figure \ref{fig:carbon}.  The posterior corresponding to the M4 mode firmly excludes the equilibrium value of $K_{\rm eq,2} \approx 0.7$, indicating that the carbon-rich model atmosphere considered is out of chemical equilibrium. The posterior corresponding to the M1 mode is only marginally consistent with the equilibrium value.

\section{Discussion}
\label{sect:discussion}

\subsection{Comparison to previous work}

\subsubsection{\cite{greene16}}

\cite{greene16} did not perform an information content analysis, but they did study mock retrievals across several exoplanet types (see their Tables 1 to 3) and JWST modes (see their Table 4), both in emission and transmission.  Six molecules were explicitly considered in the mock retrievals: CO, CO$_2$, H$_2$O, CH$_4$, NH$_3$ and N$_2$.  For transmission spectra, the transit chord was assumed to be isothermal.  The cloud model consists of a cloud top pressure (for gray clouds) and a power-law prescription (for non-gray clouds consisting of small particles).  A key finding of \cite{greene16} is: ``$\lambda=1$--2.5 $\mu$m transmission spectra will often constrain the major molecular constituents of clear solar-composition atmospheres well."

The fourth rows of Figures 7 and 8 of \cite{greene16} show mock retrievals for a warm Neptune (700 K) and warm sub-Neptune (600 K), respectively, with clouds and solar metallicity.  It is worth noting that \cite{greene16} have fixed $X_{\rm CO_2} = 3.16 \times 10^{-11}$ and $X_{\rm CO} = 10^{-9}$ for both cases (see their Table 3).  While the comparison is imperfect, the M2 mode (0.97--1.84 $\mu$m) may be compared to the NIRISS mode (1--2.5 $\mu$m) considered by \cite{greene16}.  Our RvP analysis in Figure \ref{fig:M2_RvP} (Appendix) suggests that CO (with ${\cal R}^2=0.039$ for the M2 mode) is undetectable across the wavelength range of NIRISS, which is consistent with the unconstrained posterior distributions of $X_{\rm CO}$ obtained by \cite{greene16} in the fourth rows of their Figures 7 and 8.  Since the contributions of CO and CO$_2$ to C/O are negligible in both cases, we have
\begin{equation}
\mbox{C/O} = \frac{X_{\rm CO}+X_{\rm CO_2} + X_{\rm CH_4}}{X_{\rm CO}+2X_{\rm CO_2} + X_{\rm H_2O}} \approx \frac{X_{\rm CH_4}}{X_{\rm H_2O}}.
\end{equation}
This explains why the posterior distributions of C/O associated with the 1--2.5 $\mu$m versus 1--5 $\mu$m retrievals are similar in the fourth rows of Figures 7 and 8 of \cite{greene16}.

As a further check, the third row of Figure 6 of \cite{greene16}, which describes a mock retrieval for a hot Jupiter ($X_{\rm CO} \sim 10^{-4}$), shows an unconstrained posterior distribution of $X_{\rm CO}$ associated with 1--2.5 $\mu$m.  However, the posterior distribution of $X_{\rm CO}$ associated with 1--5 $\mu$m is bounded on both sides, consistent with the findings of the current study.

\subsubsection{\cite{howe17}}
\label{subsect:howe17}

\cite{howe17} traded model sophistication for a broad exploration of the JWST modes of the NIRcam, NIRISS, NIRSpec and MIRI instruments (see their Table 1), including the proposal of a set of observing programs for hot Jupiters (see their Table 2).  Mock atmospheric retrievals are performed using a Markov Chain Monte Carlo code.  Their Table 3 lists the 11 hot Jupiters considered in their study.  Figure 7 of \cite{howe17} shows examples of calculations of Jacobians with respect to the metallicity, temperature and pressure.  Even though \cite{howe17} suggest the use of Jacobians to diagnose cloud properties, they ultimately do not explore this option in their study.  For reasons of computational feasibility, \cite{howe17} opted for a 3-parameter model that explores the temperature (of the isothermal transit chord), metallicity and cloud top pressure (or equivalently, a constant cloud opacity).

\cite{howe17} remarked that, ``For our simple forward model, the instrument that consistently gives the most information is the NIRISS G700XD mode," which corresponds to a wavelength range of 0.6--2.8 $\mu$m.  At face value, the statement about the NIRISS G700XD mode appears to be at odds with the conclusions of the current study that the blue modes of NIRSpec are suboptimal for constraining the elemental abundances and C/O (see Section \ref{subsect:compare}).  The solution to this conundrum lies in the assumption of chemical equilibrium made by \cite{howe17}.  In chemical equilibrium, knowledge of the elemental abundances, temperature and pressure allows one to fully specify all of the molecular abundances.  Equivalently, one can back out the elemental abundances if the temperature, pressure and only a subset of the molecular abundances are known.  

For example, at a given temperature and pressure one can infer O/H given \textit{only} $X_{\rm H_2O}$ if chemical equilibrium is assumed (e.g., \citealt{heng18}).  It bypasses the need to detect CO or CO$_2$, which are generally needed, in a chemical disequilibrium situation, for accurately inferring O/H.  If the ratios of O/H to the other elemental abundances are further assumed to take on their solar values, then the metallicity may be inferred as a single number (e.g., \citealt{heng18}).  Otherwise, the metallicity is generally a set of numbers given by the different elemental abundances.  The inferred information content of the 0.6--2.8 $\mu$m mode hinges on accepting these assumptions.

\subsubsection{\cite{bl17}}

\cite{bl17} used an approach to information content (IC) analysis that is similar to that of \cite{howe17}, which is based on computing Jacobians.  Their model explorations are based on a WASP-62b-like gas giant, where the main parameters are the temperature, C/O and metallicity.  It is unclear if the C/H or O/H  has a fixed (solar) ratio to the other elemental abundances.  The cloud model follows that of \cite{greene16}.  Key conclusions from \cite{bl17} include: ``A single observation with NIRISS always yields the highest IC content spectra with the tightest constraints, regardless of temperature, C/O, [M/H], cloud effects or precision."  As elucidated in Section \ref{subsect:howe17}, this conclusion hinges on the assumption of chemical equilibrium.  The temperature range considered by \cite{bl17} (600--1800 K) crosses the transition where chemical equilibrium starts to break down at low temperatures.

\subsubsection{\cite{nm20}}

In a recent study, \cite{nm20} assessed the random forest technique for atmospheric retrieval. They compared several retrievals using both random forests and the traditional nested-sampling method. They also added the extension of a likelihood function to the forest to produce posteriors that match the nested-sampling retrievals. The close agreement between their extended random forest and the nested-sampling posteriors is unsurprising as the same likelihood function is used in both. The agreement implies consistency and not necessarily veracity. 

In their comparisons, \cite{nm20} show some discrepancies between the standard random forest and the nested-sampling retrievals. In an improvement on the implementation of \cite{mn18}, we have upgraded the trees in the forest to predict the entire set of parameters in the given leaf, as opposed to taking the average value of each leaf (as described in Section \ref{subsect:rf}). This gives a more accurate sampling of the posterior. This upgrade is not included in the standard random forest used in \cite{nm20}, and could account for the discrepancies in their Figure 13. 

\cite{nm20} also discuss the issue in \cite{cobb09}, who showed an example where the forest predicts an overconfident, incorrect value of ammonia at the prior minimum in a mock retrieval. As discussed in Section 4.4 and Figure A4 of \cite{fisher20}, this effect arises from a limitation of the training set used and not because of the random forest. Specifically, because spectroscopic line list data needed to compute the ammonia opacity did not exist above 1500 K, the ammonia mixing ratio was artificially set to $10^{-13}$ when the temperature crossed this threshold. \cite{fisher20} showed that this artefact was also detected using the nested-sampling method. In other words, \cite{cobb09} succeeded in identifying the limitation of the training set, but drew the wrong conclusion from their findings.

In Section 4 of \cite{nm20}, it is suggested that the forest cannot be used for a retrieval with many parameters, claiming that: ``A Random Forest retrieval with $n$ free parameters appears to require $\gtrsim 10^n$ models for an adequate training set." There is in fact no explicit rule for the size of the training set, which will likely depend on many variables such as the relationships between the parameters, the prior ranges, the resolution of the spectra, etc. We found no issues in the current study when using our 11-parameter model on both the WFC3- and JWST-like spectra. One can see from the predicted vs real plots that the forest's performance is quite reasonable given the degeneracies one expects from multiple parameters.

\subsection{Recommendations for JWST observing proposals}
\label{subsect:compare}

\begin{figure*}%[!th]
\begin{center}
\vspace{-0.1in}
\includegraphics[width=2\columnwidth]{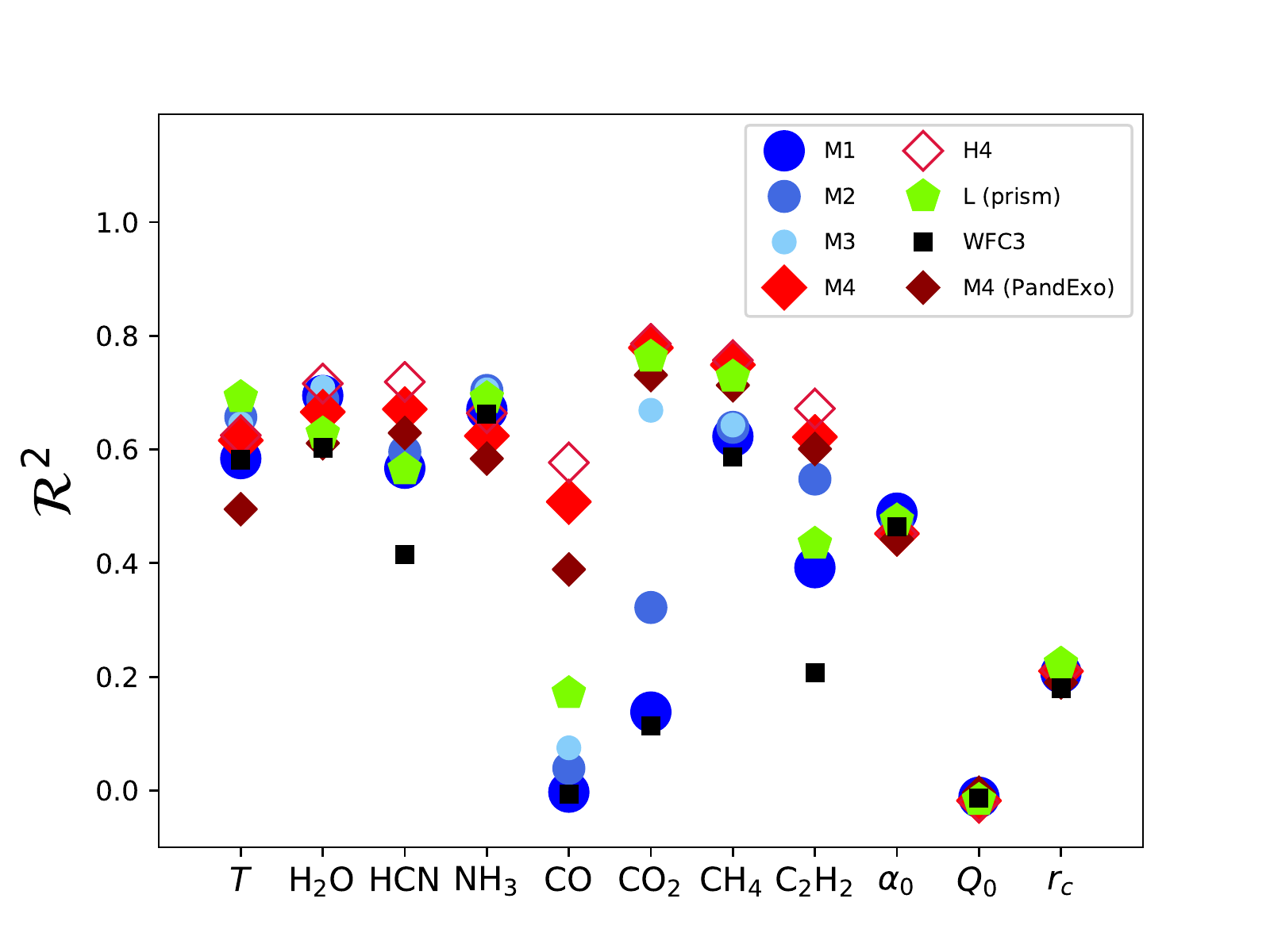}
\end{center}
%\vspace{-0.1in}
\caption{Constraining power of various JWST NIRSpec observing modes as quantified by the coefficient of determination (${\cal R}^2$).  See Table 1 for an explanation of the modes and wavelength coverage.  Zero and perfect predictability correspond to ${\cal R}^2=0$ and ${\cal R}^2=1$, respectively.  For comparison, the WFC3 channel (0.8--1.7 $\mu$m) of HST is included.  The H4 mode covers the same wavelength range as the M4 mode, but at a higher resolution of $\sim 2700$.}
%\vspace{-0.05in}
\label{fig:compare}
\end{figure*}

In Figure \ref{fig:compare}, we consolidate all of our findings into a summary plot that quantifies the predictive power of every JWST NIRSpec mode considered in the current study.  Several key points arise from inspecting Figure \ref{fig:compare}.
\begin{itemize}

    \item The three bluest medium-resolution modes (M1, M2 and M3) are essentially blind to CO (${\cal R}^2 \approx 0$), implying that the derived elemental abundances of carbon and oxygen may be inaccurate if CO is a major constituent, data are only available in these modes ($\lesssim 1.8$ $\mu$m) and the atmospheric abundances are out of chemical equilibrium.
    
    \item All of the modes do equally well at constraining $\alpha_0$ (which subsumes the cloud abundance) and the cloud particle size (which is constrained by the slope of the spectral continuum), but do equally poorly at identifying cloud composition via constraining the \textit{change in slope} of the spectral continuum.
    
    \item Perhaps the most surprising finding is that the M4 mode (2.87--5.10 $\mu$m) out-performs the low-resolution ($\sim 100$) prism mode (0.6--5.3 $\mu$m) on the ability to constrain every parameter except for the temperature and ammonia abundance. Both modes constrain the cloud properties equally well (or poorly).  In the trade-off between spectral resolution (by a factor $\sim 10$) and wavelength coverage, the former triumphs.
    
    \item While increasing the resolution from $\sim 100$ to $\sim 1000$ enhances the constraining power substantially, a further increase of resolution to $\sim 2700$, corresponding to the high-resolution modes of JWST NIRSpec, adds diminishing value.  We demonstrate this by performing a RvP analysis of the M4 mode with a resolution of $\sim 2700$, which we label as ``H4" in Figure \ref{fig:compare}. On average, the ${\cal R}^2$ value increases by 0.026 or about 5.3\% across the 11 parameters. The biggest improvement in ${\cal R}^2$ is associated with CO: from 0.508 to 0.577 (increase of 13.6\%).

\end{itemize}

\begin{figure*}%[!th]
\begin{center}
\vspace{0.1in}
\includegraphics[width=\textwidth]{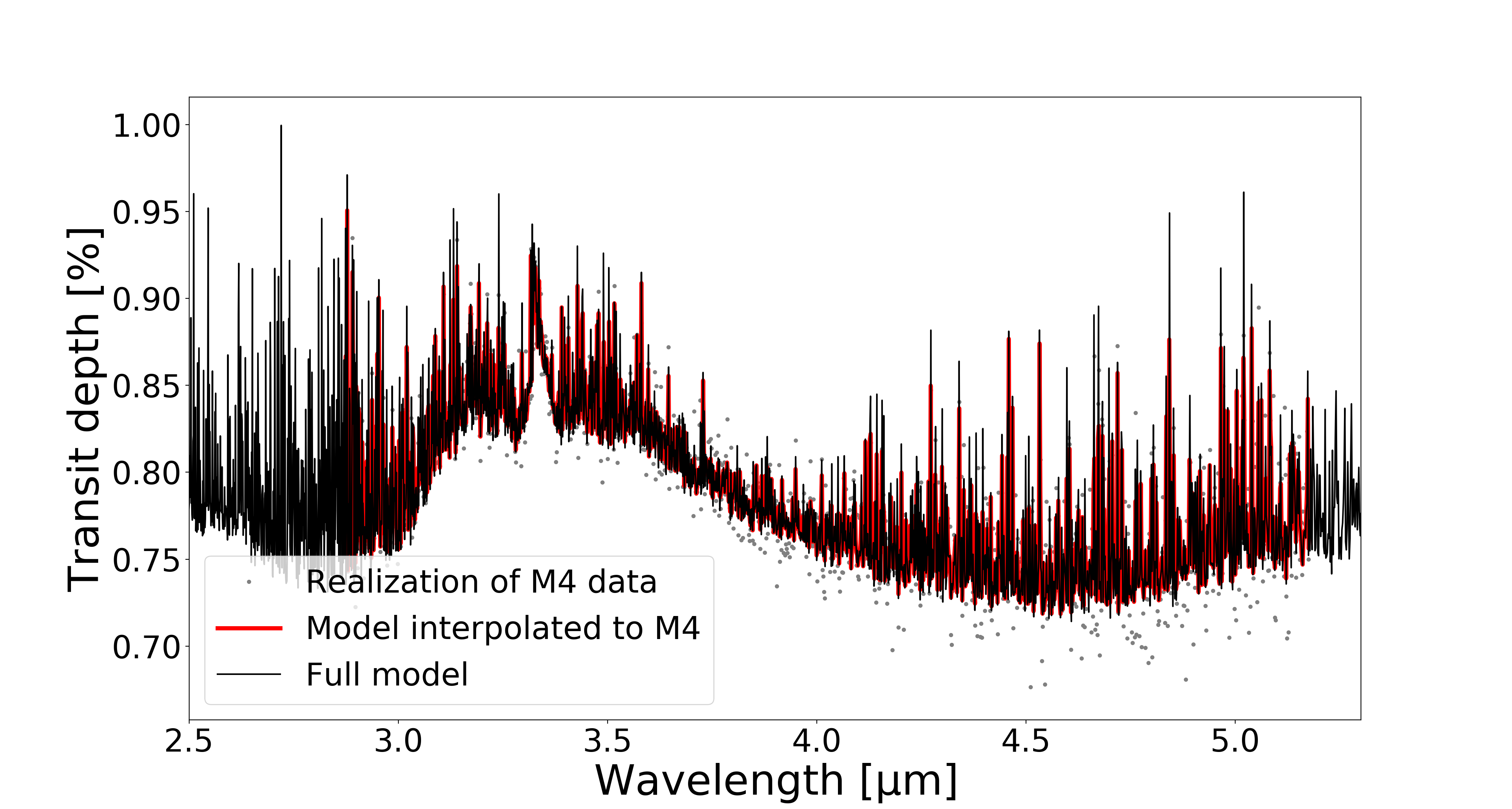}
\end{center}
\vspace{-0.1in}
\caption{A sample model spectrum from the training set, computed at a resolution of 3000 over the full wavelength range (black) and interpolated onto the wavelength grid of the M4 mode (NIRSpec G395M; red). The gray points denote a single realization of the model with errors sampled from the noise model as computed by \texttt{PandExo}.}
\vspace{-0.1in}
\label{fig:pandexo}
\end{figure*}

In this study, we have adopted a fiducial noise model in which every spectral value is sampled with an uncertainty of 20 ppm, which is the optimistic theoretical noise floor of JWST \citep{beichman14}.  As a sensitivity test, we perform another set of calculations using \texttt{PandExo} \citep{batalha17} to simulate a Bright Object Time-Series observation of GJ 436b and to obtain a more realistic noise model for application to the M4 mode.  The noise model is simulated by assuming a single transit time series of GJ 436b with the G395M grating and the sub2048 subarray read-out mode. The standard deviation as predicted by \texttt{PandExo} varies between 200 and 550 ppm over the M4 wavelength range. Our model spectra that serve as training data are subsequently interpolated onto the wavelength grid simulated by \texttt{PandExo}.  An example model spectrum after interpolation and addition of noise is shown in Figure \ref{fig:pandexo}.

Figure \ref{fig:compare} shows the constraining power for various model parameters obtained using the different modes, including M4 with the realistic noise model obtained with \texttt{PandExo}. Despite the fact that the noise of the realistic model is $\sim 10$ to $20 \times$ higher than initially assumed, the qualitative conclusions remain unchanged: the M4 mode's ability (or inability) to constrain the 11 parameters of the model are similar to when 20 ppm uncertainties are assumed. The exception is CO, where the ${\cal R}^2$ value drops from 0.508 to 0.389.  However, the ${\cal R}^2$ value associated with CO$_2$ remains high: 0.731 versus 0.779.

Overall, we recommend that the medium-resolution M4 mode be used as it offers the most balanced portfolio of constraining power across temperature, molecular abundances and cloud properties.  If the goal is to constrain these parameters accurately in order to infer the elemental abundances and C/O without assuming chemical equilibrium, the medium-resolution M4 mode is sufficient; the corresponding high-resolution mode is unnecessary.

\begin{acknowledgments}
We acknowledge financial support from the Swiss National Science Foundation, the European Research Council (via a Consolidator Grant to KH; grant number 771620), the PlanetS National Center of Competence in Research (NCCR), the Center for Space and Habitability (CSH) and the Swiss-based MERAC Foundation. We are grateful to Brice-Olivier Demory for constructive discussions and advice on the manuscript.
\end{acknowledgments}

\appendix

\section{Additional Figures}
\begin{figure}[!h]
\begin{center}
%\vspace{-0.1in}
\includegraphics[width=0.8\columnwidth]{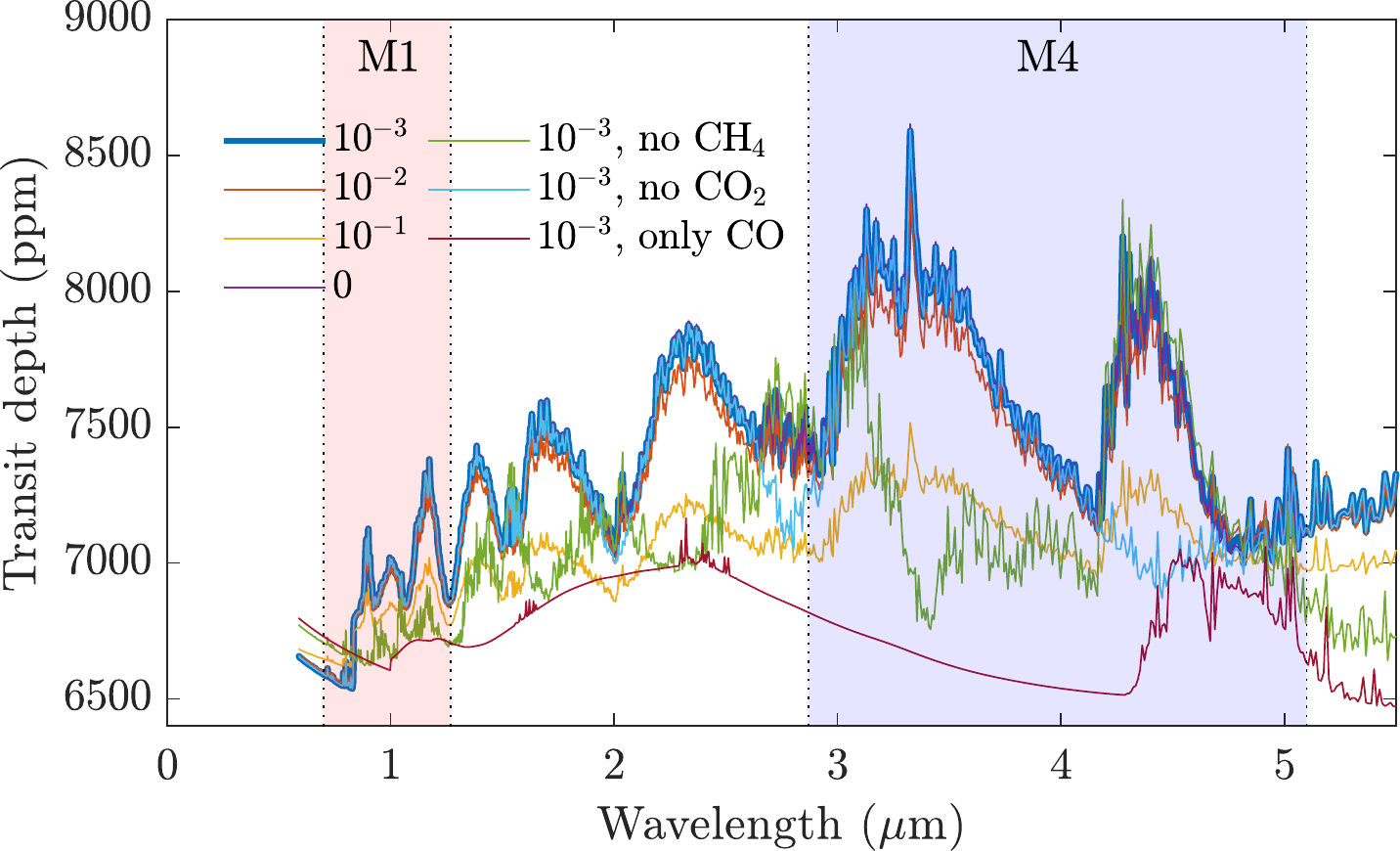}
\end{center}
\vspace{-0.1in}
\caption{Transmission spectra corresponding to the carbon-rich case study of Figure \ref{fig:carbon}, but with the CO abundance removed (labeled ``0") or varied from $10^{-3}$ (its default value) to $10^{-1}$. Three additional curves with CO only, CH$_4$ removed and CO$_2$ removed are included.}
%\vspace{-0.1in}
\label{fig:explain}
\end{figure}

Figure \ref{fig:explain} shows various transmission spectra associated with the carbon-rich case study of Section \ref{subsect:posteriors}. It is apparent that the transmission spectra with $X_{\rm CO}=0, 10^{-3}$ and $10^{-2}$ are very similar. The similarity of these spectra is due to the spectral lines of CO being masked by those of CH$_4$ and CO$_2$ at the chosen abundances ($X_{\rm CO}=X_{\rm CH_4}=10^{-3}$, $X_{\rm CO_2}=10^{-4}$).  The transmission spectrum with $X_{\rm CO}=0.1$ is markedly different only because CO is so abundant that it changes the mean molecular mass---and hence the pressure scale height---significantly. 

\begin{figure}[!h]
\begin{center}
%\vspace{-0.1in}
\includegraphics[width=\columnwidth]{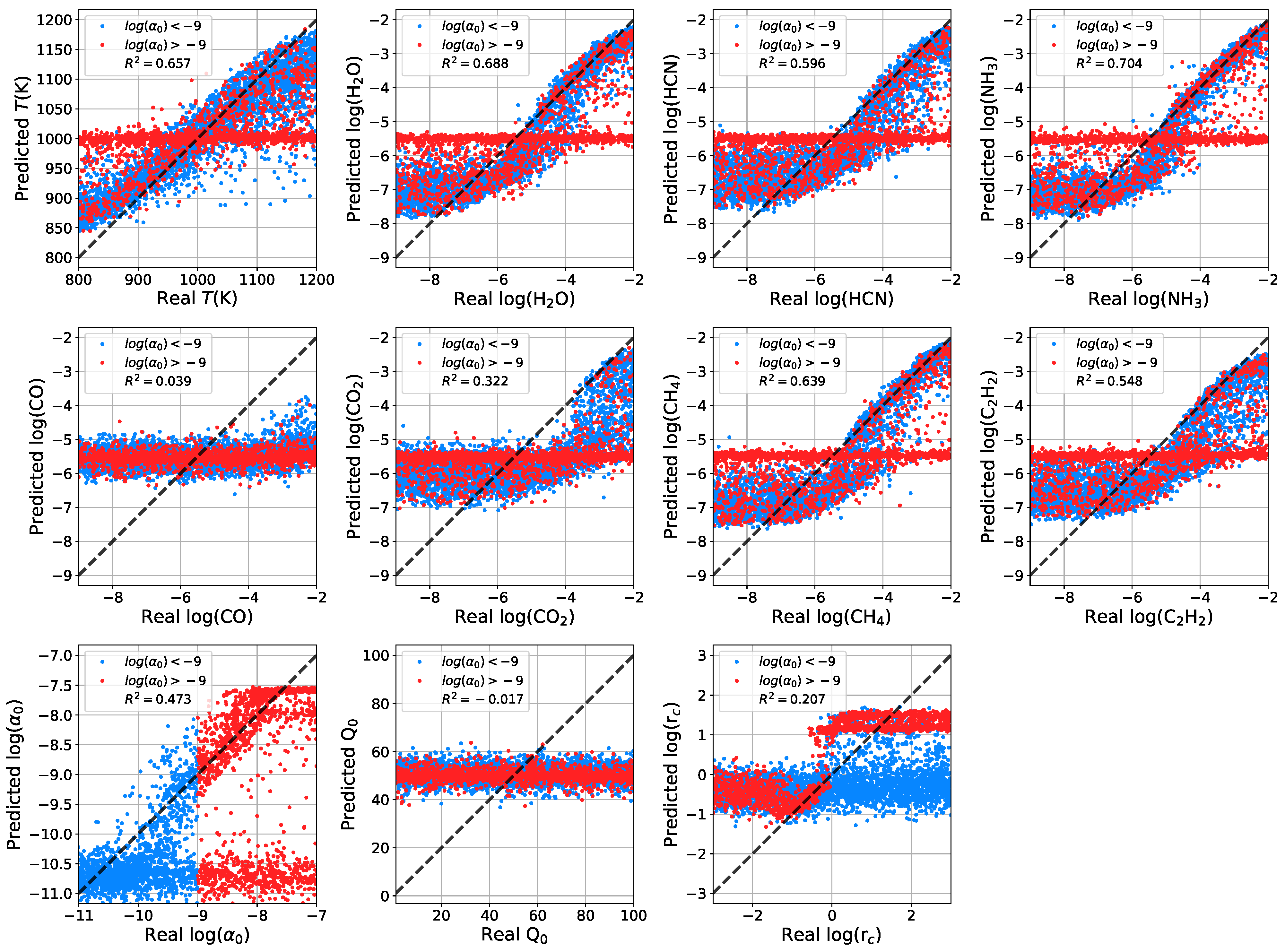}
\end{center}
\vspace{-0.1in}
\caption{Same as Figure \ref{fig:rvp}, but for the M2 mode.}
%\vspace{-0.1in}
\label{fig:M2_RvP}
\end{figure}

\begin{figure}[!h]
\begin{center}
%\vspace{-0.1in}
\includegraphics[width=0.8\columnwidth]{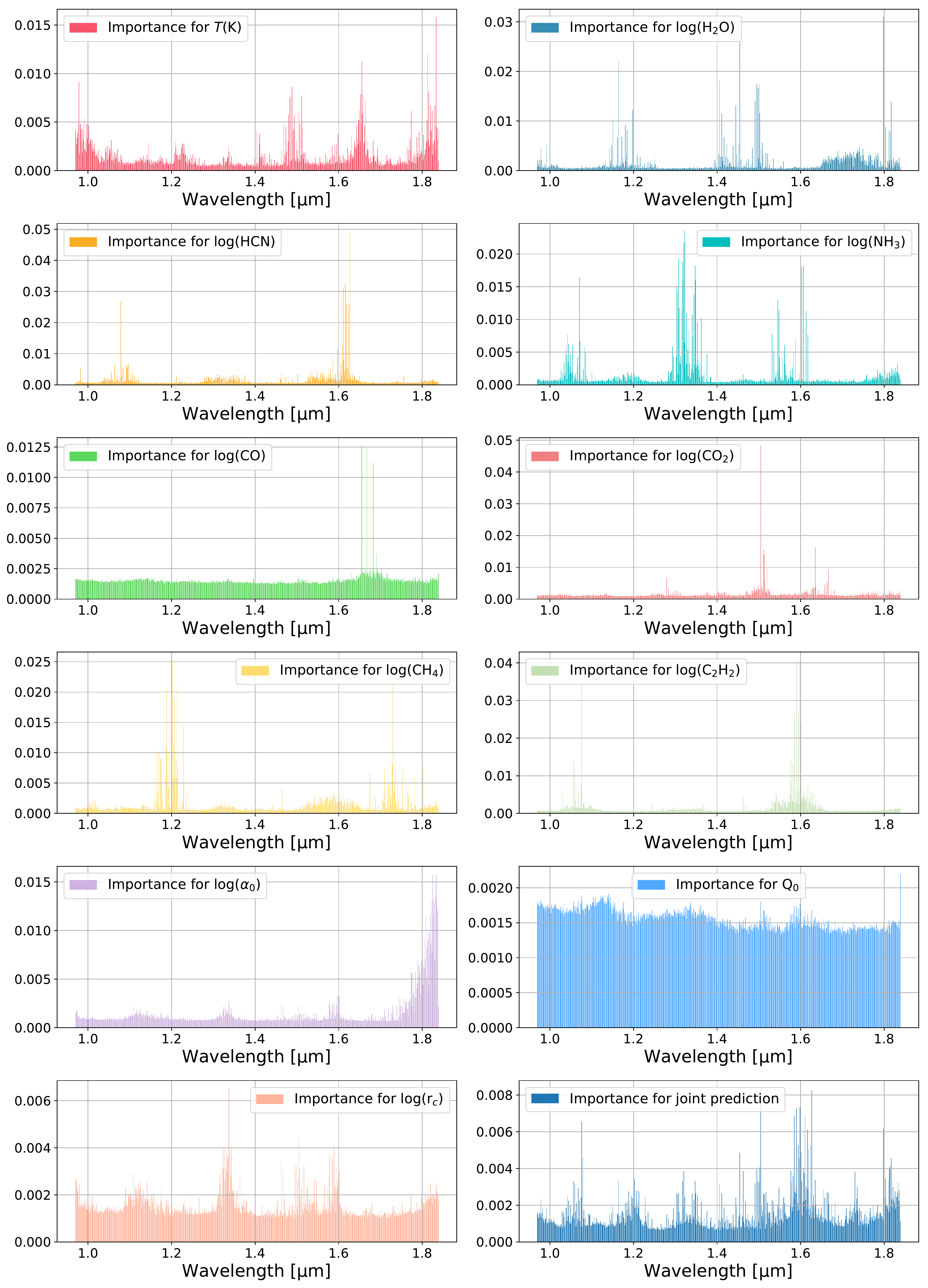}
\end{center}
\vspace{-0.1in}
\caption{Same as Figure \ref{fig:feature}, but for the M2 mode.}
%\vspace{-0.1in}
\label{fig:M2_feature}
\end{figure}

\begin{figure}[!h]
\begin{center}
%\vspace{-0.1in}
\includegraphics[width=\columnwidth]{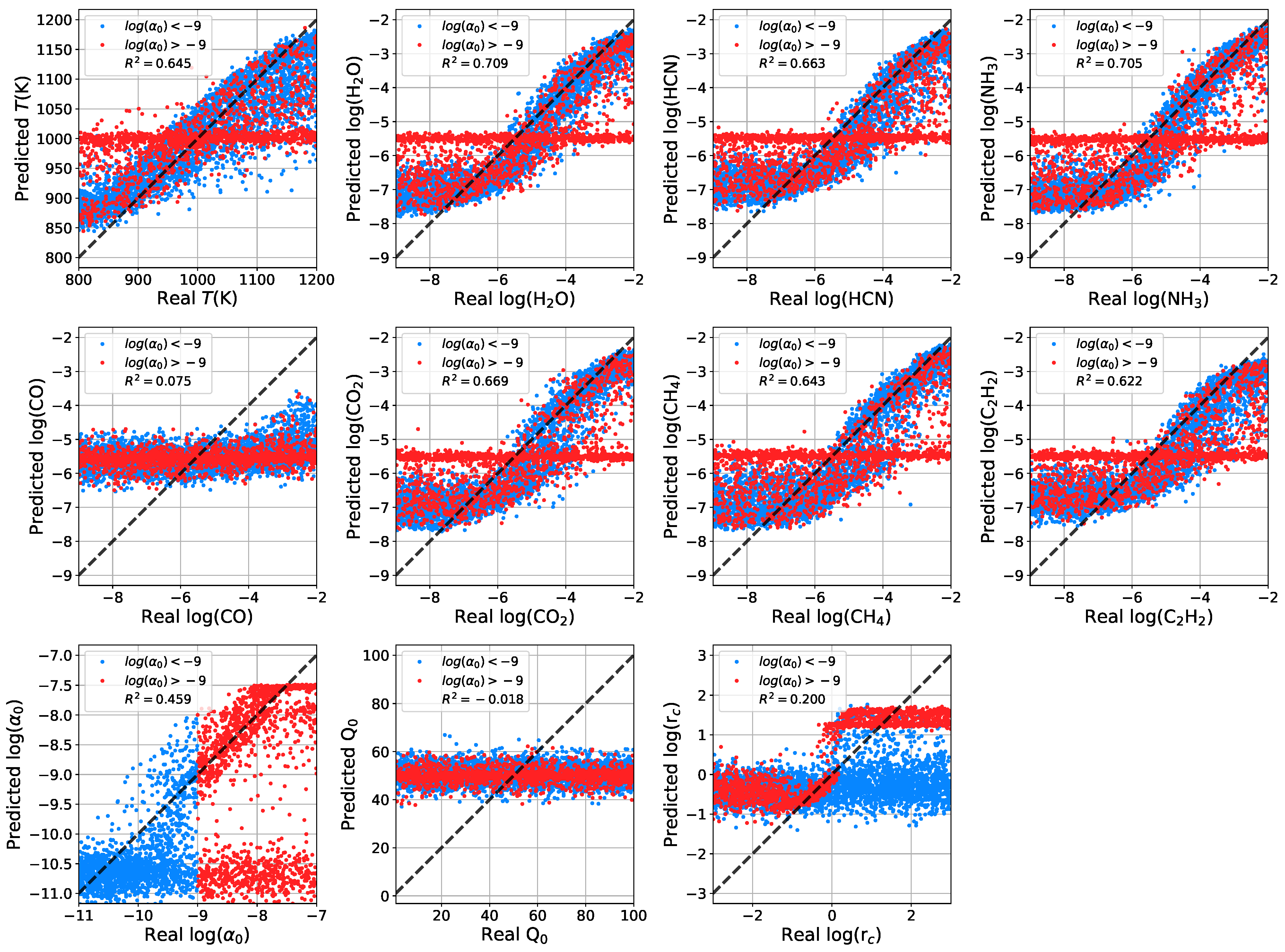}
\end{center}
\vspace{-0.1in}
\caption{Same as Figure \ref{fig:rvp}, but for the M3 mode.}
%\vspace{-0.1in}
\label{fig:M3_RvP}
\end{figure}

\begin{figure}[!h]
\begin{center}
%\vspace{-0.1in}
\includegraphics[width=0.8\columnwidth]{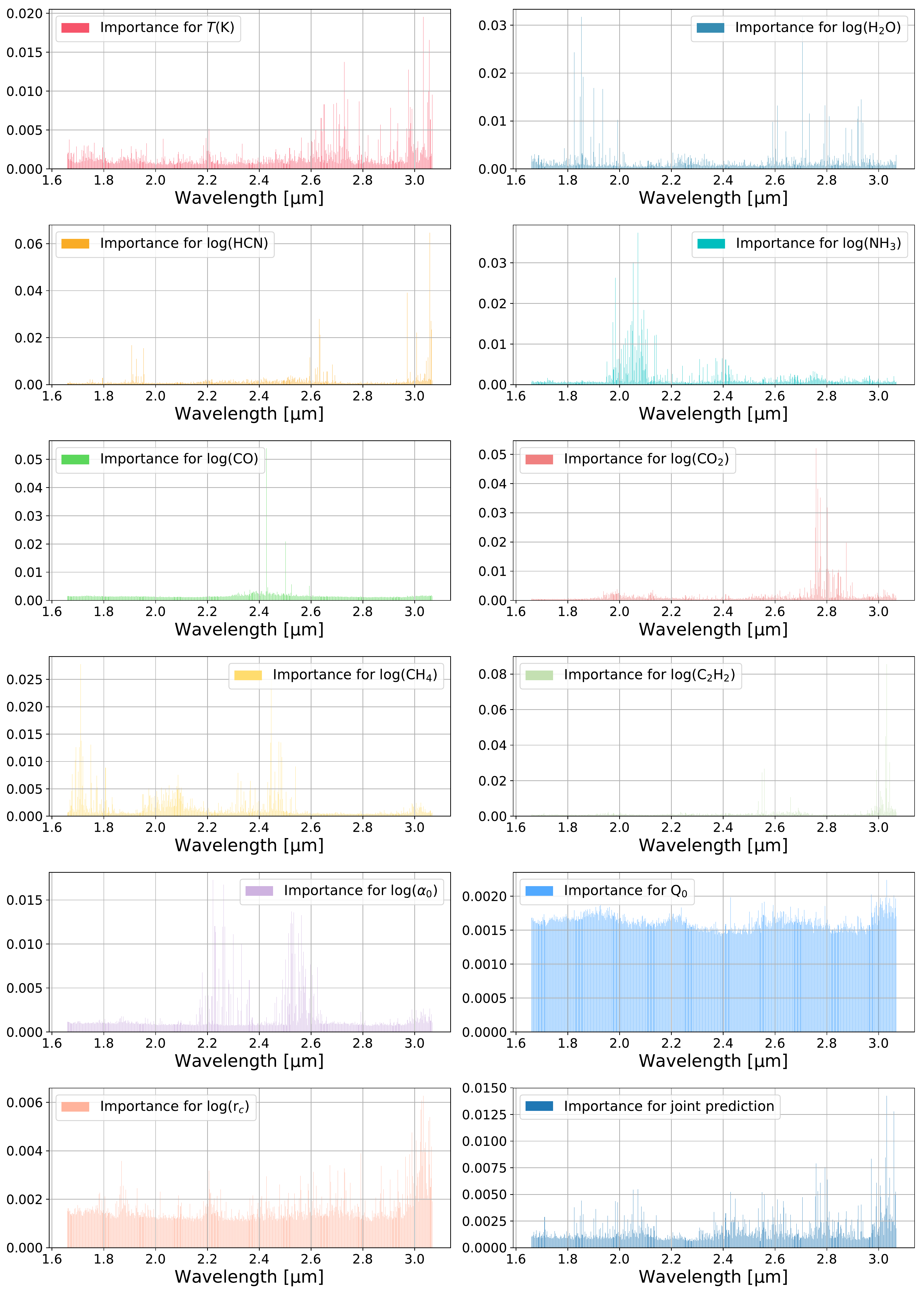}
\end{center}
\vspace{-0.1in}
\caption{Same as Figure \ref{fig:feature}, but for the M3 mode.}
%\vspace{-0.1in}
\label{fig:M3_feature}
\end{figure}

\begin{figure}[!h]
\begin{center}
%\vspace{-0.1in}
\includegraphics[width=\columnwidth]{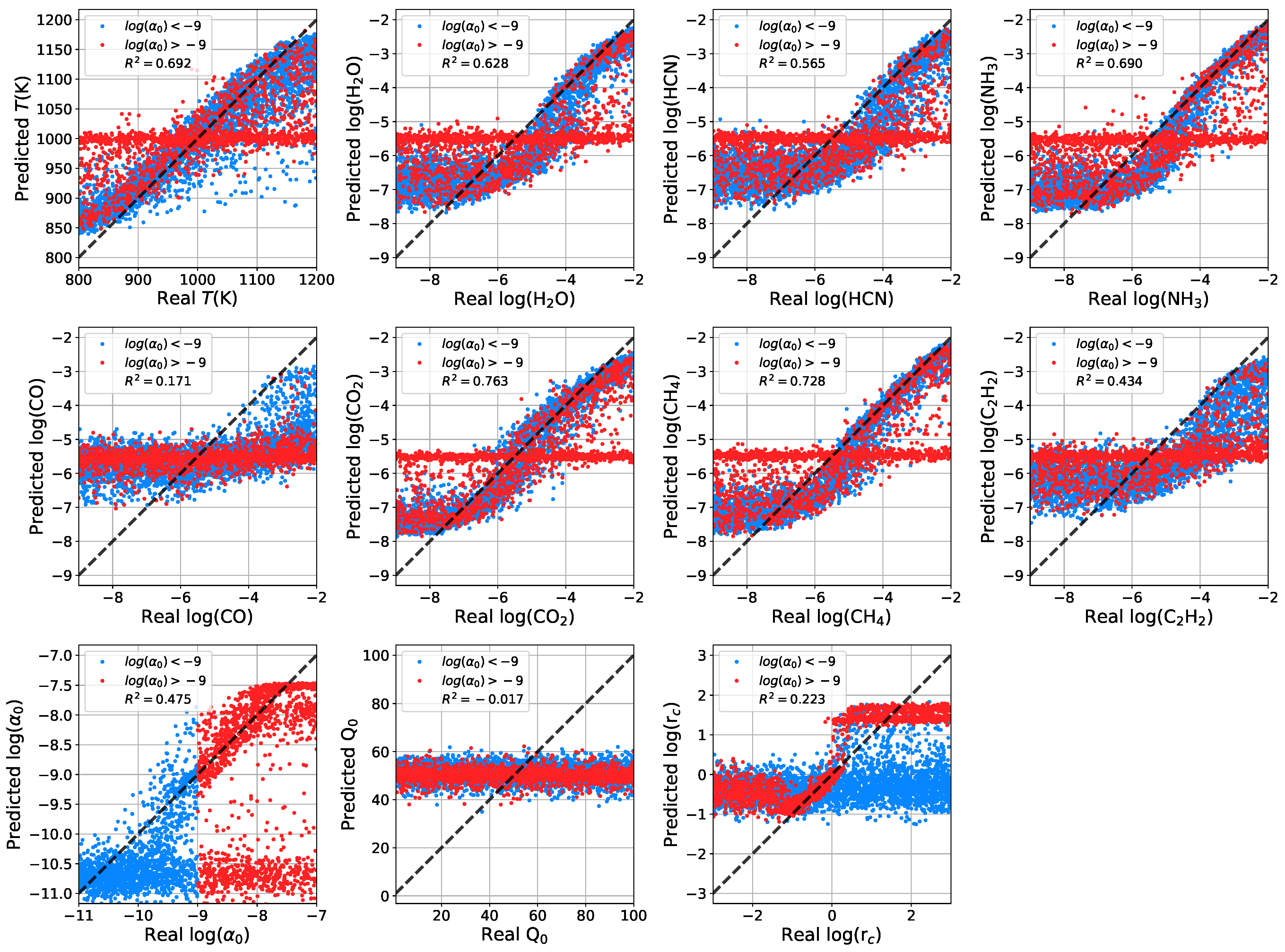}
\end{center}
\vspace{-0.1in}
\caption{Same as Figure \ref{fig:rvp}, but for the L mode.}
%\vspace{-0.1in}
\label{fig:L_RvP}
\end{figure}

\begin{figure}[!h]
\begin{center}
%\vspace{-0.1in}
\includegraphics[width=0.8\columnwidth]{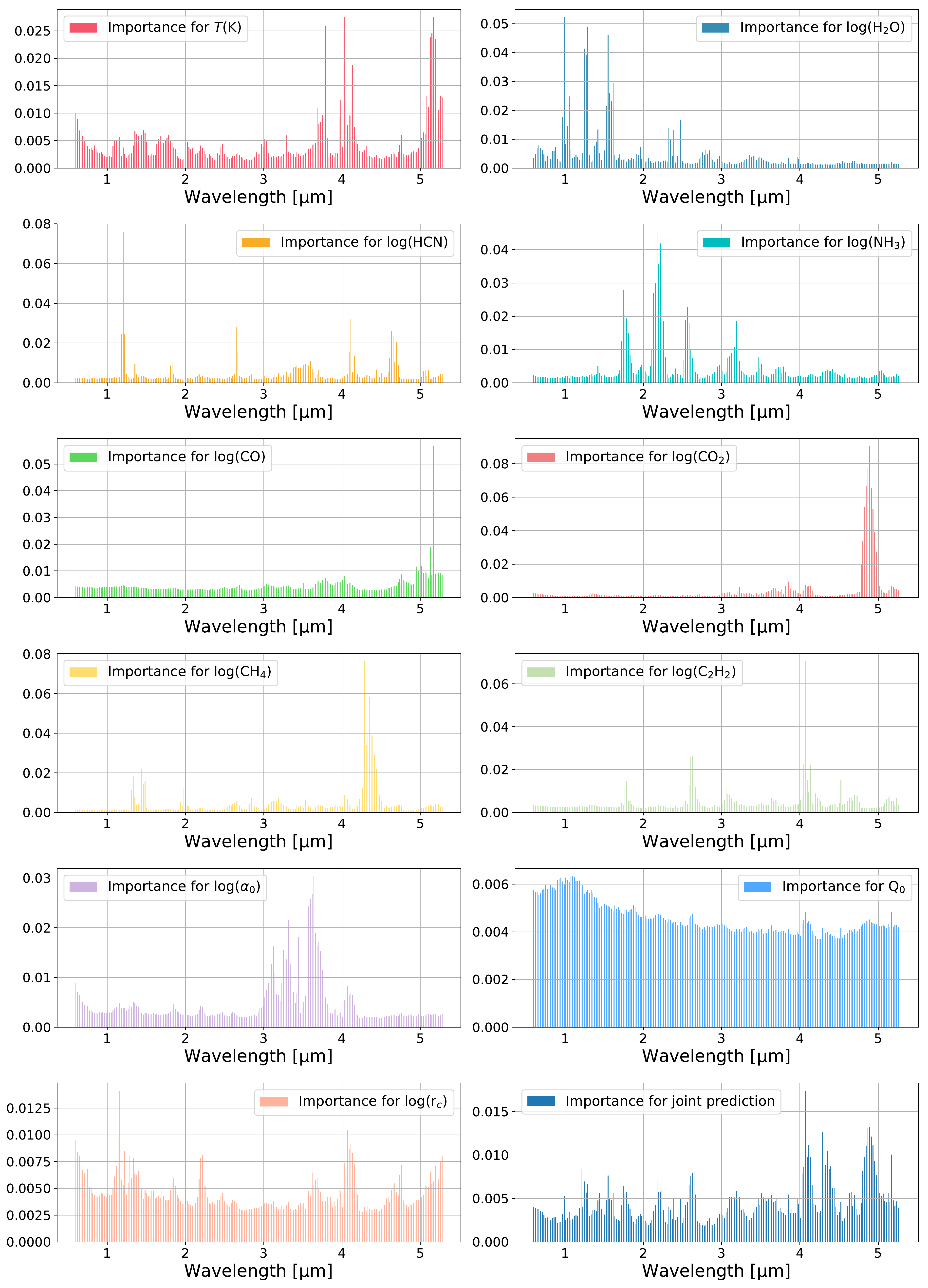}
\end{center}
\vspace{-0.1in}
\caption{Same as Figure \ref{fig:feature}, but for the L mode.}
%\vspace{-0.1in}
\label{fig:L_feature}
\end{figure}

For completeness, we include in Figures \ref{fig:M2_RvP} to \ref{fig:L_feature} the RvP and feature importance plots of the M2, M3 and L modes.

\label{lastpage}


\begin{thebibliography}{99}

\bibitem[Asplund et al.(2009)]{asplund09} Asplund, M., Grevesse, N., Sauval, A.J., \& Scott, P. \ 2009, ARAA, 47, 481

\bibitem[Barber et al.(2006)]{barber06} Barber, R.J., Tennyson, J., Harris, G.J., \& Tolchenov, R.N. \ 2006, MNRAS, 368, 1087

\bibitem[Barber et al.(2014)]{barber14} Barber, R.J., Strange, J.K., Hill, C., et al. \ 2014, MNRAS, 437, 1828

\bibitem[Barstow et al.(2015)]{barstow15} Barstow, J.K., Aigrain, S., Irwin, P.G.J., Kendrew, S., \& Fletcher, L.N. \ 2015, MNRAS, 448, 2546

\bibitem[Batalha \& Line(2017)]{bl17} Batalha, N.E., \& Line, M.R. \ 2017, AJ, 153, 151

\bibitem[Batalha et al.(2017)]{batalha17} Batalha, N.E., \& Mandell, A. \& Pontoppidan, K., et al. \ 2017, PASP, 129, 064501 

\bibitem[Bean et al.(2018)]{bean18} Bean, J.L., Stevenson, K.B., Batalha, N.M., et al. \ 2018, PASP, 130, 114402

\bibitem[Beichman et al.(2014)]{beichman14} Beichman, C., Benneke, B., Knutson, H., et al.\ 2014, PASP, 126, 1134

\bibitem[Benneke \& Seager(2012)]{bs12} Benneke, B., \& Seager, S. \ 2012, ApJ, 753, 100

\bibitem[Bower et al.(2019)]{bower19} Bower, D.J., Kitzmann, D., Wolf, A.S., et al. \ 2019, A\&A, 631, A103

\bibitem[Breiman(2001)]{breiman01} Breiman, L. \ 2001, Machine Learning, 45, 5

\bibitem[Brown(2001)]{brown01} Brown, T.M. \ 2001, ApJ, 553, 1006

\bibitem[Burrows \& Sharp(1999)]{bs99} Burrows, A., \& Sharp, C.M. \ 1999, ApJ, 512, 843

\bibitem[Charbonneau et al.(2008)]{char08} Charbonneau, D., Knutson, H.A., Barman, T., et al. \ 2008, ApJ, 686, 1341

\bibitem[Cobb et al.(2019)]{cobb09} Cobb, A.D., Himes, M.D., Soboczenski, F., et al \ 2019, AJ, 158, 33

\bibitem[Criminisi et al.(2011)]{crimi11} Criminisi, A., Shotton, J., \& Konukoglu, E. \ 2011, Microsoft Research technical report, TR-2011-114

\bibitem[Crossfield et al.(2016)]{crossfield16} Crossfield, I.J.M., Ciardi, D.R., Petigura, E.A., et al. \ 2016, ApJS, 226, 7

\bibitem[Cushing et al.(2009)]{cushing09} Cushing, M.C., Looper, D., Burgasser, Adam J., et al. \ 2009, ApJ, 696, 986

\bibitem[Dragomir et al.(2019)]{drago19} Dragomir, D., Teske, J., G\"{u}nther, M.N., et al. \ 2019, ApJ, 875, L7

\bibitem[Drummond et al.(2019)]{drummond19} Drummond, B., Carter, A.L., H\'{e}brard, E., Mayne, N.J., Sing, D.K., Evans, T.M., \& Goyal, J. \ 2019, MNRAS, 486, 1123

\bibitem[Esposito et al.(2019)]{esposito19} Esposito, M., Armstrong, D.J., Gandolfi, D., et al. \ 2019, A\&A, 62, A165

\bibitem[Feroz \& Hobson(2008)]{feroz08} Feroz, F., \& Hobson, M.P. \ 2008, MNRAS, 384, 449

\bibitem[Feroz et al.(2009)]{feroz09} Feroz, F., Hobson, M.P., \& Bridges, M. \ 2009, MNRAS, 398, 1601

\bibitem[Feroz et al.(2013)]{feroz13} Feroz, F., Hobson, M.P., Cameron, E., Pettitt, A.N. \ 2013, arXiv:1306.2144

\bibitem[Fisher \& Heng(2018)]{fh18} Fisher, C., \& Heng, K. \ 2018, MNRAS, 481, 4698

\bibitem[Fisher et al.(2020)]{fisher20} Fisher, C., Hoeijmakers, H.J., Kitzmann, D., et al. \ 2020, AJ, 159, 192

\bibitem[Fortney et al.(2005)]{fortney05} Fortney, J.J., Marley, M.S., Lodders, K., Saumon, D., \& Freedman, R. \ 2005, ApJ, 627, L69

\bibitem[Fortney et al.(2006)]{fortney06} Fortney, J.J., Cooper, C.S., Showman, A.P., Marley, M.S., \& Freedman, R.S. \ 2006, ApJ, 652, 746

\bibitem[Fortney et al.(2010)]{fortney10} Fortney, J.J., Shabram, M., Showman, A.P., Lian, Y., Freedman, R.S., Marley, M.S., \& Lewis, N.K. \ 2010, ApJ, 709, 1396
\bibitem[Gordon et al.(2016)]{gordon16} Gordon, I.E., Rothman, L.S., Hill, C et al. \ 2017, Journal of Quantitative Spectroscopy and Radiative Transfer, 203, 3 

\bibitem[Gaidos et al.(2017)]{gaidos17} Gaidos, E., Kitzmann, D., \& Heng, K. \ 2017, MNRAS, 468, 3418

\bibitem[Greene et al.(2016)]{greene16} Greene, T.P., Line, M.R., Montero, C., et al. \ 2016, ApJ, 817, 17

\bibitem[Griffith(2014)]{g14} Griffith, C.A. \ 2014, Philosophical Transactions of the Royal Society A, 372, 86

\bibitem[Grimm \& Heng(2015)]{gh15} Grimm, S.L., \& Heng, K. \ 2015, ApJ, 808, 182

\bibitem[Harris et al.(2006)]{harris6} Harris, G.J., Tennyson, J., Kaminsky, B. M., Pavlenko, Ya. V., \& Jones,H. R. A. \ 2006, MNRAS, 367, 400 

\bibitem[Heng \& Lyons(2016)]{hl16} Heng, K., \& Lyons, J.R. \ 2016, ApJ, 817, 149

\bibitem[Heng \& Tsai(2016)]{ht16} Heng, K., \& Tsai, S.-M. \ 2016, ApJ, 829, 104

\bibitem[Heng \& Kitzmann(2017)]{hk17} Heng, K., \& Kitzmann, D. \ 2017, MNRAS, 470, 2972

\bibitem[Heng(2017)]{heng17} Heng, K. \ 2017, Exoplanetary Atmospheres: Theoretical Concepts and Foundations (Princeton University Press)

\bibitem[Heng(2018)]{heng18} Heng, K. \ 2018, RNAAS, 2, 128

\bibitem[Heng(2019)]{heng19} Heng, K. \ 2019, MNRAS, 490, 3378

\bibitem[Ho(1998)]{ho98} Ho, T.K. \ 1998, IEEE Transactions on Pattern Analysis and Machine Intelligence, 20, 832

\bibitem[Howe et al.(2017)]{howe17} Howe, A.R., Burrows, A., \& Deming, D. \ 2017, ApJ, 835, 96

\bibitem[Kilpatrick et al.(2018)]{kil18} Kilpatrick, B.M., Cubillos, P.E., Stevenson, K.B., et al. \ 2018, AJ, 156, 103

\bibitem[Kitzmann \& Heng(2018)]{kh18} Kitzmann, D., \& Heng, K. \ 2018, MNRAS, 475, 94

\bibitem[Kitzmann et al.(2018)]{kitzmann18} Kitzmann, D., Heng, K., Rimmer, P.B., et al. \ 2018, ApJ, 863, 183

\bibitem[Kreidberg et al.(2015)]{k15} Kreidberg, L., Line, M.R., Bean, J.L., et al. \ 2015, ApJ, 814, 66

\bibitem[Lecavelier des Etangs et al.(2008)]{lec08} Lecavelier des Etangs, A., Pont, F., Vidal-Madjar, A., \& Sing, D. \ 2008, A\&A, 481, L83

\bibitem[Lee et al.(2014)]{lee14} Lee, J.-M., Irwin, P.G.J., Fletcher, L.N., Heng, K., \& Barstow, J.K. \ 2014, ApJ, 789, 14

\bibitem[Li et. al.(2015)]{li15} Li, G., Gordon, I. E., Rothman, L. S., et al. \ 2015, ApJS, 216, 15 

\bibitem[Line et al.(2013)]{line13} Line, M.R., Wolf, A.S., Zhang, X., et al. \ 2013, ApJ, 775, 137

\bibitem[Line \& Yung(2013)]{ly13} Line, M.R., \& Yung, Y. \ 2013, ApJ, 779, 3

\bibitem[Madhusudhan \& Seager(2011)]{ms11} Madhusudhan, N., \& Seager, S. \ 2011, ApJ, 729, 41

\bibitem[Madhusudhan(2012)]{madhu12} Madhusudhan, N. \ 2012, ApJ, 758, 36

\bibitem[M\'{a}rquez-Neila et al.(2018)]{mn18} M\'{a}rquez-Neila, P., Fisher, C., Sznitman, R., \& Heng, K. \ 2018, Nature Astronomy, 2, 719

\bibitem[Mie(1908)]{mie} Mie, G. \ 1908, Annalen der Physik, 330, 377

\bibitem[Nixon \& Madhusudhan(2020)]{nm20} Nixon, M.C., \& Madhusudhan, N. \ 2020, MNRAS, in press (arXiv:2004.10755)

\bibitem[Morley et al.(2017)]{morley17} Morley, C.V., Knutson, H., Line, M., et al. \ 2017, AJ, 153, 86

\bibitem[Moses et al.(2011)]{moses11} Moses, J.I., Visscher, C., Fortney, J.J., et al. \ 2011, ApJ, 737, 15

\bibitem[Moses et al.(2013)]{moses13} Moses, J.I., Line, M.R., Visscher, C., et al. \ 2013, ApJ, 777, 34

\bibitem[Oppenheimer et al.(1998)]{opp98} Oppenheimer, B.R., Kulkarni, S.R., Matthews, K., \& van Kerkwijk, M.H. \ 1998, ApJ, 502, 932

\bibitem[Oreshenko et al.(2020)]{oreshenko20} Oreshenko, M., Kitzmann, D., M\'{a}rquez-Neila, P., et al. \ 2020, AJ, 159, 6

\bibitem[Pedregosa et al.(2011)]{pedregosa11} Pedregosa, F., Varoquaux, G., Gramfort, A., et al. \ 2011, JMLR, 12, 2825-2830

\bibitem[Petigura et al.(2013)]{petigura13} Petigura, E.A., Marcy, G.W., \& Howard, A.W. \ 2013, ApJ, 770, 69

\bibitem[Polyansky et al.(2018)]{poly18} Polyansky, O.L., Kyuberis, A.A., Zobov, N.F., et al. \ 2018, MNRAS, 480, 2597

\bibitem[Prinn \& Barshay(1977)]{pb77} Prinn, R.G., \& Barshay, S.S. \ 1977, Science, 198, 1031

\bibitem[Quinn et al.(2019)]{quinn19} Quinn, S.N., Becker, J.C., Rodriguez, J.E., et al. \ 2019, AJ, 158, 177

\bibitem[Roston \& Obaid(2005)]{ro05} Roston, G.D., \& Obaid, F.S. \ 2005, Journal of Quantitative Spectroscopy \& Radiative Transfer, 94, 255

\bibitem[Rothman et al.(1987)]{rothman87} Rothman, L.S., Gamache, R.R., Goldman, A., et al. \ 1987, Applied Optics, 26, 4058

\bibitem[Rothman et al.(1992)]{rothman92} Rothman, L.S., Gamache, R.R., Tipping, R.H., et al. \ 1992, Journal of Quantitative Spectroscopy \& Radiative Transfer, 48, 469

\bibitem[Rothman et al.(1998)]{rothman98} Rothman, L.S., Rinsland, C.P., Goldman, A., et al. \ 1996, Journal of Quantitative Spectroscopy \& Radiative Transfer, 60, 665

\bibitem[Rothman et al.(2003)]{rothman03} Rothman, L.S., Barbe, A., Benner, D.C., et al. \ 2003, Journal of Quantitative Spectroscopy \& Radiative Transfer, 82, 5

\bibitem[Rothman et al.(2005)]{rothman05} Rothman, L.S., Jacquemar, D., Barbe, A., et al. \ 2005, Journal of Quantitative Spectroscopy \& Radiative Transfer, 96, 139

\bibitem[Rothman et al.(2009)]{rothman09} Rothman, L.S., Gordon, I.E., Barber, R.J., et al. \ 2010, Journal of Quantitative Spectroscopy \& Radiative Transfer, 111, 2139

\bibitem[Rothman et al.(2010)]{rothman10} Rothman, L.S., Gordon, I.E., Barbe, A., et al. \ 2009, Journal of Quantitative Spectroscopy \& Radiative Transfer, 110, 533

\bibitem[Rothman et al.(2013)]{rothman13} Rothman, L.S., Gordon, I.E., Babikov, Y., et al. \ 2013, Journal of Quantitative Spectroscopy \& Radiative Transfer, 130, 4

\bibitem[Sisson et al.(2019)]{sisson19} Sisson, S.A., Fan, Y., \& Beaumont, M.A. \ 2019, Handbook of Approximate Bayesian Computation (CRC Press)

\bibitem[Skilling(2006)]{skilling06} Skilling, J. \ 2006, Bayesian Analysis, 1, 833

\bibitem[Stevenson et al.(2016)]{stevenson16} Stevenson, K.B., Lewis, N.K., \& Bean, J.L., et al. \ 2016, PASP, 128, 094401

\bibitem[Sudarsky et al.(2003)]{sudarsky03} Sudarsky, D., Burrows, A., \& Hubeny, I. \ 2003, ApJ, 588, 1121

%\bibitem[Sneep \& Ubachs(2005)]{su05} Sneep, M., \& Ubachs, W. \ 2005, Journal of Quantitative Spectroscopy \& Radiative Transfer, 92, 293

\bibitem[Tennyson \& Yurchenko(2017)]{ty17} Tennyson, J., \& Yurchenko, S.N. \ 2017, Molecular Astrophysics, 8, 1

\bibitem[Torres et al.(2008)]{torres08} Torres, G., Winn, J.N., \& Holman, M.J. \ 2008, ApJ, 677, 1324 

\bibitem[Trifonov et al.(2019)]{trifonov19} Trifonov, T., Rybizki, J., \& K\"{u}rster, M. \ 2019, A\&A, 622, L7

\bibitem[Trotta(2008)]{trotta08} Trotta, R. \ 2008, Contemporary Physics, 49, 71

\bibitem[Tsai et al.(2017)]{tsai17} Tsai, S.-M., Lyons, J.R., Grosheintz, L., et al. \ 2017, ApJS, 228, 20

\bibitem[Venot et al.(2014)]{venot14} Venot, O., Ag\'{u}ndez, M., Selsis, F., Tessenyi, M., \& Iro, N. \ 2014, A\&A, 562, A51

\bibitem[von Braun et al.(2012)]{vb12} von Braun, K., Boyajian, T.S., Kane, S.R., et al. \ 2012, ApJ, 753, 171

\bibitem[Waldman et al.(2015)]{waldmann15} Waldmann, I.P., Tinetti, G., Rocchetto, M., et al. \ 2015, ApJ, 802, 107

\bibitem[Waldmann(2016)]{waldmann16} Waldmann, I.P. \ 2016, ApJ, 820, 107

\bibitem[Yurchenko et al.(2011)]{y11} Yurchenko, S.N., Barber, R.J., \& Tennyson, J. \ 2011, MNRAS, 413, 1828

\bibitem[Yurchenko et al.(2013)]{y13} Yurchenko, S.N., Tennyson, J., Barber, R.J., \& Thiel, W. \ 2013, Journal of Molecular Spectroscopy, 291, 69

\bibitem[Yurchenko \& Tennyson(2014)]{yt14} Yurchenko, S.N., \& Tennyson, J. \ 2014, MNRAS, 440, 1649

\bibitem[Yurchenko et al.(2018)]{y18} Yurchenko, S.N., Al-Refaie, A.F., \& Tennyson, J. \ 2018, A\&A, 614, A131

%\bibitem[Zaghloul(2007)]{z07} Zaghloul, M.R. \ 2007, MNRAS, 375, 1043

\end{thebibliography}
\end{document}